\documentclass[12pt]{article}

\usepackage{paper2e}
\usepackage{mydefs2e}
\usepackage{xspace}
\usepackage{epsfig}

\newcommand{\Sla}[1]%
{\kern0.12em{\raise.15ex\hbox{$/$}\kern-.74em #1}}

\renewcommand{\Box}{\,\raisebox{-.45pt}{\drawsquare{7}{0.6}}\,}
\newcommand{\grad}{\vec\nabla}

\newcommand{\MP}{M_{\rm Pl}}

\renewcommand{\d}{\partial}
\newcommand{\cd}[1]{\nabla_{\!#1}\,}

\newcommand{\dddot}[1]{\stackrel{\rm ...}{#1}}

\begin{document}

\begin{titlepage}
\preprint{\bf HUTP-06/A0006, UTAP-551}

\title{Spontaneous Lorentz Breaking at High Energies}

\author{Hsin-Chia Cheng$\,^{\rm a,b}$,
Markus A. Luty$\,^{\rm a,c,d}$,
Shinji Mukohyama$\,^{\rm a,e}$\ and\ Jesse Thaler$\,^{\rm a}$}

\address{$^{\rm a}$Jefferson Laboratory of Physics, Harvard University,\\
Cambridge, Massachusetts 02138}

\address{$^{\rm b}$Department of Physics, University of California, Davis, California 95616}

\address{$^{\rm c}$Physics Department, Boston University,
Boston, Massachusetts 02215}

\address{$^{\rm d}$Physics Department, University of Maryland,
College Park, Maryland 20742}

\address{$^{\rm e}$Department of Physics and 
Research Center for the Early Universe,\\
The University of Tokyo, Tokyo 113-0033, Japan}

\begin{abstract}
Theories that spontaneously break Lorentz invariance also violate
diffeomorphism symmetries, implying the existence of extra
degrees of freedom and modifications of gravity.
In the minimal model (``ghost condensation'') with only a single
extra degree of freedom at low energies, the scale of Lorentz violation
cannot be larger than about $M \sim 100\GeV$
due to an infrared instability in the gravity sector.
We show that Lorentz symmetry can be broken at much higher scales in a
non-minimal theory with additional degrees of freedom,
in particular if Lorentz symmetry is broken by the vacuum expectation
value of a vector field.
This theory can be constructed by gauging ghost
condensation, giving a systematic effective field theory
description that allows us to estimate the size of all physical effects.
We show that nonlinear effects become important for gravitational fields
with strength $\sqrt{\Phi} \gsim g$, where $g$ is the gauge coupling, and
we argue that the nonlinear dynamics is free from singularities.
We then analyze the phenomenology of the model, including nonlinear
dynamics and velocity-dependent effects. 
The strongest bounds on the gravitational sector come from either black
 hole accretion or 
direction-dependent gravitational forces, and imply that the scale of
 spontaneous Lorentz breaking is $M \lsim \mathrm{Min} (10^{12} \GeV,\, g^2 \, 10^{15} \GeV)$.
If the Lorentz breaking sector couples directly to matter, there is a spin-dependent
inverse-square law force, which has a different angular dependence from the
force mediated by the ghost condensate, providing a
distinctive signature for this class of models.
\end{abstract}

\end{titlepage}

\section{Introduction}

As the 100th anniversary of special relativity comes to a close, Lorentz invariance reigns as arguably the most important symmetry in modern physics.  A good way to understand the central role of Lorentz invariance is to consider how it could be violated.  Like baryon and lepton number in the standard model, Lorentz invariance could be an accidental symmetry of the leading interactions in quantum field theory.  Alternatively, Lorentz symmetry could be one limit of a more fundamental symmetry, just as the Galilean group is the small velocity limit of the Lorentz group.  But in the absence of experimental data, it is difficult to guess what deeper organizing principle could replace Lorentz invariance.  Indeed, the Lorentz group itself is one of many deformations of the Galilean group, and it was the experimental crisis of the ether that drove the transition from absolute time to the space-time continuum in 1905.

Here, we consider the less radical possibility that Lorentz invariance is a good symmetry at high energies, but is spontaneously broken at an energy scale $M$.  Some of the consequences of breaking  Lorentz invariance have been
extensively explored in the literature.
If the symmetry breaking sector couples to standard model fields,
there will be Lorentz-violating operators in the effective theory that
give rise to CPT violation and a variety of preferred frame effects
(see
\Ref{Colladay:1998fq,CPT1,CPT2,Bluhm:2005uj,Mattingly:2005re,Amelino-Camelia:2005qa,Vucetich:2005ra}
and references therein). 
A second aspect that has been received less attention is that spontaneous breaking
of Lorentz invariance implies the existence of new degrees of freedom
analogous to the Goldstone bosons that arise from spontaneous breaking of internal
symmetries.

The minimal model of this kind was analyzed in \Ref{Arkani-Hamed:2003uy},
which contains a single Goldstone degree of freedom.
It  can be viewed as the result of ghost condensation
(in the same way that ordinary spontaneous symmetry breaking can
be viewed as ``tachyon condensation'') and so we call the extra degree of
freedom a ``ghostone'' boson.
If the ghostone mode couples to the standard model, then it
gives rise to exotic spin- and velocity-dependent forces
\cite{Arkani-Hamed:2003uy,Arkani-Hamed:2004ar}.  In a gravitational theory, breaking Lorentz invariance immediately implies a violation of diffeomorphisms, the gauge group of Einstein gravity, and thus the ghostone mode will mix with the metric.  This is the Higgs mechanism for gravity, and is analogous to the familiar Higgs mechanism for gauge theories,
where the mixing between a gauge field and a Goldstone boson
gives rise to a massive spin-1 boson.
\Refs{Arkani-Hamed:2003uy,Arkani-Hamed:2005gu} analyzed the modification of gravity
that arises in this way, and found the limit
$M \lsim 100\GeV$ because of an infrared
gravitational instability analogous to the Jeans instability
for ordinary matter.  Therefore, spontaneous Lorentz breaking is directly tied to modifications of gravity,  and we can place strong constraints on Lorentz breaking by looking at gravitational physics.

In this paper, we consider a non-minimal model of spontaneously broken
Lorentz invariance that has three extra degrees of freedom instead of a single ghostone mode.
We will show that in this model it is possible to break Lorentz invariance
spontaneously at much higher scales, as high as $M \sim \mathrm{Min} (10^{12} \GeV,\, g^2 \, 10^{15} \GeV)$.
The model can be thought of as arising from a vector field order parameter,
and has been previously considered in the literature \cite{Will:1972,Hellings:1973,Jacobson:2001yj,Eling:2003rd,Jacobson:2004ts,Kostelecky:1989jw,Kostelecky:2003fs,Clayton:1998hv,Clayton:2001vy,Moffat:1992ud,Moffat:2002nm,Gripaios:2004ms,Bertolami:1998dn}.
There are several features of our analysis that are new.
First, we show that this model can be viewed as a gauging (in the ordinary sense)
of ghost condensation.
This considerably clarifies the nature of the effective theory,
and allows us to consistently estimate
the size of all operators in a low energy expansion.
Second, we consider the couplings of the Goldstone degrees of freedom to
matter, and show that there are new spin- and velocity-dependent forces with
a different signature from the minimal model.
Third, we show that the theory has important nonlinear corrections in strong
gravitational fields.
We find that the nonlinear effects are much less dramatic than in ghost
condensation (see \Ref{Arkani-Hamed:2005gu}), and argue that the theory is free from caustic
singularities.
Finally, we consider the modification of gravity in this model, studying observable consequences
such as velocity-dependent effects (such as gravitational \v{C}erenkov radiation), nonlinear dynamics, and the accretion of the condensate by black holes.
Our conclusion is that the model is consistent with all experimental
limits for $M \lsim \mathrm{Min} (10^{12} \GeV,\, g^2 \, 10^{15} \GeV)$, where $g$ is the gauge coupling
and other dimensionless coefficients are taken as ${\cal O}(1)$.

This paper is organized as follows.
In Sec.~\ref{sec:bottomup}, we describe the effective theory 
by directly introducing the extra degrees of freedom required by
diffeomorphism invariance.
This is the most direct route to the effective theory and the modification
of gravity.
In Sec.~\ref{sec:topdown}, we describe the same effective theory as the gauged version of
ghost condensation.
This formulation
makes it manifest that this is a theory of \emph{spontaneous} breaking of
Lorentz symmetry, and also makes it easy to understand the general
power counting and the construction of the nonlinear theory.
In Sec.~\ref{sec:togravity}, we couple the gauged ghost condensate to gravity
and perform a linear analysis. We find that the modification of gravity
is mild in this case and it is the form of a direction-dependent Newtonian potential.
In Sec.~\ref{sec:nonlinear}, we study the nonlinear effects. In contrast to the ungauged
ghost condensate, the would-be caustics are smoothed out by the gauge interaction
and there are no dangerous instabilities. 
In Sec.~\ref{sec:BH}, we consider gauged ghost condensate 
surrounding a black hole and calculate the rate of increase of the black
hole mass due to accretion of the condensate. 
In Sec.~\ref{sec:Cerenkov}, we compute the energy loss
due to the ether \v{C}erenkov radiation for a source moving faster than the
ghostone mode. In Sec.~\ref{sec:sminteractions}, we study various Lorentz-violating effects and their
constraints if the gauged ghost condensate couples directly to the Standard Model fields.
Sec.~\ref{sec:conclusions} contains our conclusions. Some more detailed analyses and discussion
are included in the appendices.

\section{Lorentz Violation and Diffeomorphisms}
\label{sec:bottomup}

In a non-gravitational theory, Lorentz symmetry can be broken explicitly without introducing any Goldstone degrees of freedom.  The reason is that Lorentz symmetry is a physical symmetry that arranges matter states into multiplets, so violating Lorentz symmetry will only affect the allowed couplings between matter states.%
\footnote{Of course, one can spontaneously break Lorentz symmetry in a non-gravitational theory, which would give rise to Goldstone modes.  The point is that in the absence of gravity, Lorentz violation can be explicit, whereas in a gravitational theory, Lorentz violation must be spontaneous.}
As we will see, however, breaking Lorentz invariance implies a breaking of diffeomorphism invariance, the gauge symmetry of Einstein gravity. Since a gauge symmetry is a redundancy of description rather than a physical symmetry, breaking diffeomorphisms necessarily introduces additional degrees of freedom.  The key point is that these new modes nonlinearly realize diffeomorphisms (as well as nonlinearly realizing Lorentz transformations), and therefore their couplings to gravity are uniquely determined by diffeomorphism invariance.

In this section, we show how to construct the effective theory below the Lorentz breaking scale by explicitly adding these degrees of freedom to the Lagrangian. We will also see that different ways of nonlinearly realizing diffeomorphisms give rise to different low energy effective theories.  We review the case of ghost condensation which has only one new degree of freedom and introduce ``gauged ghost condensation'' which has three new degrees of freedom.  We find that the mixing between gravity and the gauged ghost condensate is mild, which explains why the scale of spontaneous Lorentz breaking can be much higher in this theory. Other interesting generalizations are also possible, see for example \Refs{Rubakov:2004eb,Dubovsky:2004sg,Bluhm:2004ep}.

The simplest way to break Lorentz invariance is to include the time
component of a vector field in the Lagrangian:
\beq[lorentzviolangrange]
\Delta \mathcal{L} = J^0.
\eeq
The arguments below are easily generalized to components of higher
tensors.
\Eq{lorentzviolangrange} breaks Lorentz symmetry down to $SO(3)$ rotations,
and also breaks diffeomorphism symmetry.  Moreover, there is no way to couple $\Delta \mathcal{L}$ to the metric
$g_{\mu\nu}$ to restore diffeomorphisms.  To see this, we work in the linearized theory obtained by
expanding the metric about flat space (where Lorentz invariance has an unambiguous meaning)
\beq
g_{\mu\nu} = \eta_{\mu\nu} + h_{\mu\nu}.
\eeq
Under the infinitesimal diffeomorphism generated by $x^\mu \rightarrow x^\mu - \xi^\mu$, the metric and $J^\mu$ transform as
\beq
\de h_{\mu\nu} &= \d_\mu \xi_\nu + \d_\nu \xi_\mu,
\\
\de J^\mu &=  \xi^\rho \d_\rho J^\mu - \d_\rho \xi^\mu J^\rho.
\eeq
We have
\beq
\de (\sqrt{-g} \, J^0) = -\sqrt{-g} \, \d_\mu \xi^0  J^\mu,
\eeq
which makes it clear that there is no way to introduce factors of $h_{\mu\nu}$ to make $\Delta \mathcal{L} $ invariant under diffeomorphisms.

However, we can formally restore diffeomorphism invariance by introducing
a scalar field $\pi$ that transforms as
\beq
\de \pi = \xi_0.
\eeq
This is the degree of freedom in the ``ghost condensate'' model
of \Ref{Arkani-Hamed:2003uy}, and $\pi$ naturally has dimensions of length.
The combination
\beq[lorentzviolangrangepi]
\De\scr{L} = \sqrt{-g}\, (J^0 + \d_\mu \pi J^\mu)
\eeq
is fully diffeomorphism invariant.%
\footnote{In fact, \Eq{lorentzviolangrangepi} is invariant in full
nonlinear Einstein gravity.}
Note that diffeomorphism invariance is realized nonlinearly,
since the $\pi$ field transforms inhomogeneously.

Because $\pi$ is a real propagating degree of freedom, we need to introduce a kinetic term for $\pi$, but diffeomorphism
invariance requires the kinetic term to depend on $h_{\mu\nu}$.
Specifically, the time kinetic term
\beq[pikinetic]
\scr{L}_{\pi\,{\rm kinetic}} = +\sfrac 12 M^4 (\dot \pi - \sfrac 12 h_{00})^2
\eeq
is diffeomorphically invariant in the linearized theory.
Note that the sign is fixed by the requirement that $\pi$ fluctuations
have positive energy.
\Eq{pikinetic} shows that diffeomorphism invariance requires mixing between the new degree of freedom and gravity.
In a gauge where $\pi \equiv 0$, \Eq{pikinetic} becomes a ``wrong sign''
mass term for $h_{00}$, suggesting an oscillatory Newtonian potential at distances
larger than the inverse ``mass'' $r_c \sim \MP / M^2$.
A more careful analysis shows that the potential is indeed modified at
distances larger than $r_c$, but only on a much larger time scale
$t_c \sim \MP^2 / M^3$ \cite{Arkani-Hamed:2003uy}.
The theory is unstable on time scales of order $t_c$, and demanding that
this is longer than the present age of the universe gives $M \lsim 10\MeV$, though
nonlinear effects slightly relax this bound to $M \lsim 100\GeV$
\cite{Arkani-Hamed:2005gu}.

The strong bound on $M$ arises because the way we have restored diffeomorphism
invariance allows a graviton ``mass''.%
\footnote{Strictly speaking, only the Newtonian potential gets a mass; the propagating spin-2 gravitons are still massless.}
To find a theory that allows larger values of $M$, 
we look for a way of restoring diffeomorphism invariance
that does not necessitate a graviton mass term.
To do this, note that
\beq
\de [\sqrt{-g}\, (1 + \sfrac 12 h_{00}) J^0 ]
= \sqrt{-g}\, \d_i \xi^0 J^i,
\eeq
where the sum over $i$ does not include any additional minus signs.  This shows that we can restore difformorphism invariance by introducing three
fields $a_i$ transforming as
\beq[atrans]
\de a_i = \d_i \xi_0.
\eeq
Note that $a_i$ defined in this way is naturally dimensionless.
Because these fields transform as the gradient of a diffeomorphism
parameter,
the time kinetic term involves mixing with derivatives of $h_{\mu\nu}$:
\beq
\scr{L}_{a\,{\rm kinetic}} \sim  +M^2 (\dot{a}_i
- \sfrac 12 \d_i h_{00})^2.
\eeq
It is easy to see that all diffeomorphism-invariant combinations of $a_i$
and $h_{\mu\nu}$ involve derivatives acting on the metric, 
so modifications of Newton's law are suppressed.
We will analyze this theory in more detail starting in Sec.~\ref{sec:topdown},
and we will see that this construction allows much larger values for the scale
$M$ where Lorentz symmetry is broken.

To see the connection between the above theory and gauging ghost condensation, note that the
diffeomorphism transformation \Eq{atrans} is very similar to an ordinary $U(1)$ gauge
transformation
\beq
\de a_\mu = \d_\mu \la.
\eeq
In fact, we can promote \Eq{atrans} to an ordinary gauge transformation by
introducing a field $\pi$ that transforms under gauge transformations
and diffeomorphisms as
\beq
\de \pi = \xi_0 - \la.
\eeq
The combination
\beq
\tilde{a}_\mu = a_\mu +  \d_\mu \pi
\eeq
is then gauge invariant and transforms under diffeomorphisms as
\beq
\de \tilde{a}_\mu = \d_\mu \xi_0.
\eeq
In ``unitary gauge'' $\pi \equiv 0$ we have $\tilde{a}_i = a_i$, so the
spatial components can be identified with the fields introduced in equation \Eq{atrans}.
The extra degree of freedom $\tilde{a}_0$ is naturally massive, since nothing
forbids the term
\beq[a0mass]
\scr{L}_{a_0\,{\rm mass}} = + \sfrac 12 M^4 (\tilde{a}_0 - \sfrac 12h_{00})^2
\eeq
which is a mass term for $a_0$ in unitary gauge.  

As we will explain in sections \ref{sec:topdown} and \ref{sec:togravity}, \Eq{a0mass} reduces to \Eq{pikinetic} when we take the $a_\mu$ gauge coupling to zero, which explains why the modification of gravity is suppressed in gauged ghost condensation.  We find that for canonically normalized fields in unitary gauge, \Eq{a0mass} becomes 
\beq
\mathcal{L}_\mathrm{mass} = \frac{M^2}{2} \left(g A^c_0 - g_c\Phi^c  \right)^2, \qquad g_c = \frac{M}{\sqrt{2}\MP},
\eeq
where $A^c_\mu$ is the canonically normalized gauge field, $\Phi^c$ is the canonically normalized Newtonian potential, and $g$ is the $A^c_\mu$ gauge coupling.  When $g$ is smaller than $g_c$, the (negative) mass term dominantly affects the Newtonian potential, yielding large modifications to gravitational potentials as in ghost condensation.  When $g$ is larger than $g_c$, the mass term acts mostly on the ``Coulomb'' potential $A^c_0$, shielding Lorentz breaking from the gravitational sector and allowing the scale $M$ in gauged ghost condensation to be raised as high as $M \lsim \mathrm{Min} (10^{12} \GeV,\, g^2 \, 10^{15} \GeV)$.  

A discussion of the decoupling limit of gauged ghost condensation which makes contact with earlier literature on Lorentz breaking appears in App.~\ref{app:decoupling}.

\section{Gauged Ghost Condensation}
\label{sec:topdown}

In this section, we will start from the ghost
condensate and construct the effective theory by gauging a global
symmetry.
This will clarify the power counting, and also 
show how the theory reduces to the ghost condensate
in a particular limit.

\subsection{A Review of Ghost Condensation}
We first review the ``ghost condensate'' model of \Ref{Arkani-Hamed:2003uy}.
This is an effective theory of a real scalar field $\phi$ with a global
shift symmetry
\beq[phishift]
\phi \rightarrow \phi + \lambda.
\eeq
The effective theory is assumed to depend on a
single scale $M$, which gives both the strength of
self-interactions of the $\phi$ field and the cutoff of the 
effective theory.
The novel feature of this sector is that it has a vacuum
(in flat spacetime)
\beq
\avg{\d_\mu \phi} = v_\mu,
\eeq
where $v_{\mu}$ is a constant timelike vector.
This spontaneously breaks Lorentz invariance by establishing a preferred
time direction.
We can then choose a Lorentz frame where
$v_0 = c$, $v_i = 0$ for $i = 1, 2, 3$.
In this frame we have
\beq
\avg{\phi} = c t.
\eeq
It is convenient to take $\phi$ to have units of time, so that $c$ is
dimensionless and can be set to 1 by rescaling the time coordinate.

Expanding about the vacuum
\beq
\phi = \avg{\phi} + \pi,
\eeq
we can construct a systematic effective field theory for the
fluctuations $\pi$ in derivatives simply by noting that
\beq[eqn:vacuum]
\bal
\d_\mu \phi &=  \de_\mu{}^0 + \partial_\mu \pi, \\
\d_\mu \d_\nu \phi &= \d_\mu \d_\nu \pi.
\eal
\eeq
The constant part of $\partial_\mu \phi$ should be kept to all orders
since there is no momentum suppression. On the other hand, 
terms with more than one derivative acting on
$\phi$ only give rise to derivatives acting on the fluctuations,
so there is a well-defined expansion at low energies and only the 
lowest order terms are important. The most general quadratic effective
Lagrangian with at most four derivatives acting on $\pi$ is
\beq[Lghostnonlinear]
\bal
\scr{L}_{\rm ghost} &= M^4 P(X) + \sfrac 12 M^2 \Bigl(
Q_1(X) (\Box \phi)^2
+ Q_2(X) \d^\mu \phi \d^\nu \phi \d_\mu \d_\nu \phi \Box\phi
\\
& \qquad\qquad\qquad\qquad\quad
+ Q_3(X) (\d^\mu \phi \d^\nu \phi \d_\mu \d_\nu \phi)^2 \Bigr)
+\cdots,
\eal\eeq
where  we have assumed a $\phi \rightarrow -\phi$ reflection symmetry, and
\beq
X = \d^\mu \phi \d_\mu \phi.
\eeq

Expanding this Lagrangian about the vacuum using \Eq{eqn:vacuum}, the leading Lagrangian is
\beq[Lghostlinear]
\!\!\!\!\!\!\!\!
\scr{L}_{\rm ghost} = \sfrac 12 M^4 \dot\pi^2
- \sfrac 12 M^2 \Bigl(
\al (\grad^2 \pi)^2
+ \be \ddot\pi \grad^2 \pi
- \ga \ddot\pi^2 \Bigr)
+ \scr{O}\left(\pi^3, (\d^3\pi)^2\right).
\eeq
Here we have rescaled the fields so that $P''(1) = \frac 14$,
which fixes the coefficient of the $\dot\pi^2$ term, and used the fact
that $P'(1)=0$ in the vacuum so that there is no tadpole term.
The coefficients of the other terms are given by
\beq
\bal
\al &= -Q_1(1), \\
\be &= 2 Q_1(1) + Q_2(1), \\
\ga &= Q_1(1) + Q_2(1) + Q_3(1). 
\eal
\eeq
We can further rescale $\pi$ and $M$ to simplify this Lagrangian,
but we will not make use of this freedom for now.
We can see that there is no normal spatial kinetic term $(\grad \pi)^2$,
which implies the special dispersion relation for $\pi$
\begin{equation}
\omega^2 = \alpha \frac{\vec{k}^4}{M^2}
\end{equation}
to leading order in $\omega$ and $k$.

\subsection{Gauging Ghost Condensation}
We now couple the theory above to a gauge field $A_\mu$ by gauging
the shift symmetry in \Eq{phishift}.
That is, we promote \Eq{phishift} to a local transformation and introduce
a gauge field to make the Lagrangian invariant.
It is convenient to define the gauge field as
$A_\mu = M a_\mu$ so that the gauge
field has dimensions of mass.
We then write the gauge transformations as
\beq
\de A_\mu = \d_\mu \chi,
\eeq
where $\chi$ is dimensionless.
The gauge transformation of $\phi$ is then
\beq
\de \phi = - \frac{1}{M} \chi.
\eeq
We have rescaled $\chi$ and $A_\mu$ to fix the coefficients in these
transformation rules.
The fields $A_\mu$ and $\phi$ must always appear in the gauge-invariant
combination
\beq
\scr{A}_\mu = A_\mu + M \d_\mu \phi,
\eeq
which can also be thought of as the gauge covariant derivative of $\phi$:
$\scr{A}_\mu = M D_\mu \phi$.

We assume that the theory violates Lorentz invariance in the vacuum via
the gauge-invariant order parameter
\beq
\avg{\scr{A}_\mu} = M v_\mu,
\eeq
where $v_\mu$ is a timelike vector of unit norm.
(This defines the scale $M$.)
In the preferred frame, the order parameter is $\avg{\scr{A}_0}$.

The coupling of the gauge field to the ghost condensate is governed
by gauge invariance.
Starting from \Eq{Lghostlinear}, we can write terms
\beq
\scr{L}_{\rm gauged\,ghost} = -\frac{1}{4 g^2} F^{\mu\nu} F_{\mu\nu}
+ \scr{L}_{\rm ghost}(\d_\mu\phi \to \scr{A}_\mu / M).
\eeq
This does not uniquely fix the leading terms, since
we can obtain gauge invariant terms by replacing
$\d_\mu\d_\nu \phi$ with either $\d_\mu \scr{A}_\nu$ or
$\d_\nu \scr{A}_\mu$.
The general linearized Lagrangian expanded
about the vacuum $\avg{\scr{A}_\mu} = M
\delta_\mu{}^0$
is then
\beq[Lgaugedghostlin]
\bal
\scr{L}_{\rm gauged\,ghost} &= -\frac{1}{4 g^2} F^{\mu\nu} F_{\mu\nu}
+ \sfrac 12  M^2 (\scr{A}_0 - M)^2
\\
& \qquad
- \sfrac 12 \al_1 (\d_i \scr{A}_i)^2
- \sfrac 12 \al_2 (\d_i {\scr{A}}_j)^2
+ \sfrac 12 \be_1 (\d_t \scr{A}_i)^2
- \sfrac 12 \be_2 (\d_i \scr{A}_0)^2
\\
& \qquad
+ \be_3 \d_t \scr{A}_0 \d_i \scr{A}_i
+ \sfrac 12 \ga (\d_t \scr{A}_0)^2 
+ \scr{O}(\d^2 \scr{A}^4)
+ \scr{O}(\scr{A}^3).
\eal
\eeq
By the power counting of the ghost condensate effective theory, the
coefficients
$\al_i, \beta_i, \ga$ are order 1, and give rise to general kinetic terms
for the vector fields. 
While the $F^{\mu\nu} F_{\mu\nu}$ term does not generate any independent
gauge kinetic terms,
for $g \ll 1$ it gives the dominant kinetic contribution to the transverse
modes.
In the limit $g \to 0$, we recover the ghost condensate theory.

\Eq{Lgaugedghostlin} contains a ``mass'' term for $\scr{A}_0$,
so we can integrate out $\scr{A}_0$ if we are interested
in energies and momenta below the scale $M$.
To see this explicitly, we compute the dispersion relation for the
scalar modes.
We parametrize them by
\beq
A_0 = M + a,
\qquad
A_i = \d_i \si,
\qquad
\pi.
\eeq
Choosing ``unitary gauge'' $\pi \equiv 0$, the quadratic Lagrangian
in momentum space is
\beq
\scr{L} = \frac{1}{2g^2} \pmatrix{\si & a \cr} K \pmatrix{\si \cr a \cr},
\eeq
where
\beq
K = 
\pmatrix{(1 + g^2 \be_1) \om^2 \vec{k}^2 - g^2 \alpha \vec{k}^4 &
i (1 - g^2 \be_3) \om \vec{k}^2 \cr
-i (1 - g^2 \be_3) \om \vec{k}^2 &
(1 - g^2 \be_2) \vec{k}^2 + g^2 \ga \om^2 + g^2 M^2 \cr},
\eeq
and $\alpha \equiv \alpha_1 + \alpha_2$. The dispersion relation for the scalar modes is obtained by
setting the determinant of the kinetic matrix to zero.
We find two scalar modes, one with dispersion relation
\beq[goldstonedispersion]
\om^2 = \frac{\al g^2}{1 + g^2 \be_1} \vec{k}^2
+ \left( \frac{1 - g^2 \be_3}{1 + g^2 \be_1} \right)^2
\frac{\al \vec{k}^4}{M^2}
+ \scr{O}(\vec{k}^6 / M^2),
\eeq
and one with
\beq
\om^2 = -\frac{M^2}{\ga} +\scr{O}(\vec{k}^2).
\label{eqn:heavy-mode}
\eeq
The first mode is gapless and corresponds roughly to the longitudinal mode $\si$.
For sufficiently small $\vec{k}$ we can neglect the quartic term in the
dispersion relation, so the gapless modes travel with constant speed in the
preferred frame.
For $g \ll 1$, modes with $|\vec{k}| \ll g M$ have speed
\beq
c_s = \sqrt{\alpha} \, g \ll 1.
\eeq 
Gapless modes with $|\vec{k}| \gg g M$ have a quartic dispersion relation,
just as the Goldstone boson $\pi$ in the (ungauged) ghost condensate.
In this way, this theory approaches the theory of \Ref{Arkani-Hamed:2003uy}
as $g \to 0$.
In the opposite limit $g \gg 1$ the speed is $\scr{O}(1)$ for the gapless
mode with $|\vec{k}| \ll M$, so the quartic term is never important within
the regime of validity of the effective theory.

The second mode has an energy gap of order $M$, and is therefore not
a mode that is accurately described in the effective theory.
We therefore write the
effective theory below the scale $M$ by integrating out the field $A_0$,
which has an order-1 overlap with the massive mode.
We can do this without fixing any gauge, and we obtain the effective
Lagrangian
\beq[Leff]
\scr{L}_{\rm eff} = \frac{1}{2 g_t^2} (\d_t \vec{\scr{A}})^2
- \frac{1}{4 g_s^2} F_{ij}^2
- \sfrac 12 \al (\grad \cdot \vec{\scr{A}})^2
+ \scr{O}(1/M^2),
\eeq
where
\beq
\frac{1}{g_t^2} = \frac{1}{g^2} + \be_1,
\qquad
\frac{1}{g_s^2} = \frac{1}{g^2} + \al_2.
\eeq
We have dropped $1/M^2$ terms in \Eq{Leff},
which means that we have neglected
the $\vec{k}^4$ term
in the dispersion relation for the massless scalar mode.
For $g \ll 1$, this is only 
justified if we restrict attention to momenta satisfying
$|\vec{k}| \ll g M$.\footnote{The $\vec{k}^4$ correction can be obtained
by keeping the  term
$$
\De\scr{L} = -\frac{1}{2g^4 M^2}
(\partial_t \d_i \scr{A}_i)^2,
$$
which is enhanced for small $g$.}

What has happened is that the Goldstone mode $\pi$ has been ``eaten''
by the gauge field $A_\mu$ and can be viewed as the
longitudinal mode of the gauge field $\scr{A}_\mu$.
However, unlike in the Lorentz-invariant Higgs mechanism, 
only $\scr{A}_0$ gets a mass term and 
the modes parametrized by $\scr{A}_i$ remain massless.
In a sense, the Lorentz-violating Higgs mechanism 
shields ``electric'' interactions ($\scr{A}_0$) while
preserving ``magnetic'' ones ($\scr{A}_i$).
We will explore this interesting feature when we consider
coupling the gauged ghost condensate to the standard model
in Sec.~\ref{sec:sminteractions}.

\section{Coupling the Gauged Ghost Condensate to Gravity}
\label{sec:togravity}
We now analyze the coupling of the gauged ghost condensate to gravity.
We will see that the modification of gravity is much less dramatic
than in ghost condensation.
One aspect of this was already discussed in Sec.~\ref{sec:topdown}, namely that
the vector modes mix with gravity only via derivative terms, and therefore
do not modify the static Newtonian potential, at least in the
preferred frame.
Another aspect comes from the result of the previous section,
that the gauged ghost condensate has a dispersion
relation $\om^2 \simeq c_s \vec{k}^2$ for small $\vec{k}$,
so long wavelength fluctuations are stable.
Coupling to gravity can give only additional terms suppressed
by $1/\MP^2$, so we do not expect any instabilities in the presence
of gravity.
This is in contrast to ghost condensation, where the theory without
gravity has a dispersion relation $\om^2 \sim \vec{k}^4 / M^2$,
and gravity gives a correction $\De \om^2 \sim - M^2 \vec{k}^2 / \MP^2$
that generates an instability at long wavelengths.

Since we start with a completely generally covariant description of
the theory, the coupling to gravity is determined simply by promoting
the metric to a dynamical degree of freedom and replacing spacetime
derivatives $\d_\mu$ by covariant derivatives $\nabla_\mu$.
For the full nonlinear theory, this is
\beq\bal
\scr{L} = \scr{L}_{\rm EH} + \sqrt{-g} \biggl[ &
-\frac 1{4g^2} g^{\mu\nu} g^{\rho\si} F_{\mu\rho} F_{\nu\si}
+ \sfrac 14  M^2 (g^{\mu\nu} \scr{A}_\mu \scr{A}_\nu - M^2)^2
\\
& + \sfrac 12 \al_1 (g^{\mu\nu} \nabla_\mu \scr{A}_\nu)^2
+ \cdots \biggr].
\eal\eeq
where $\scr{L}_{\rm EH}$ is the Einstein-Hilbert action.

\subsection{Linear Theory}
Expanding about flat spacetime
\beq
g_{\mu\nu} = \eta_{\mu\nu} + h_{\mu\nu},
\eeq
the quadratic terms in the Lagrangian are 
\beq[Lgaugedgravghostlin]
\scr{L} &= \scr{L}_{\rm EH} 
-\frac{1}{4 g^2} F^{\mu\nu} F_{\mu\nu}
+ \sfrac 12  M^2 (\scr{A}_0 - \sfrac 12 M h_{00} - M)^2
+ \cdots
\eeq
The weak gauging of gravity affects the dispersion relation for the
scalar modes only by terms suppressed by $1/\MP$, so we can still
integrate out the massive mode $A_0$ as before.
The leading quadratic terms in the effective Lagrangian for small $g$
is then
\beq[gaugedgraveffghostL]
\scr{L}_{\rm eff} = \scr{L}_{\rm EH}
+ \frac{1}{2 g_t^2} (\cd{t} \scr{A}_i)^2
- \frac{1}{4 g_s^2} F_{ij}^2
- \sfrac 12 \al (\d_i \scr{A}_i)^2
+ \cdots.
\eeq
Here
\beq
\cd{t} \scr{A}_i = \d_t \scr{A}_i
- \frac{M}{\sqrt{2}\MP} \d_i \Phi^c,
\eeq
where $\Phi^c = \sqrt{2}\MP \Phi = \MP h_{00}/\sqrt{2}$ is the 
canonically normalized Newtonian potential.
Note that the $(\cd{t} \scr{A}_i)^2$ term contains mixing between
$\scr{A}_i$ and $h_{00}$.

The effective Lagrangian in \Eq{gaugedgraveffghostL} is fully
invariant under gauge transformations and diffeomorphisms:
\beq[gaugedtransformlaws]
\bal
\de \pi &= -\xi_0 - \frac{\chi}{M},\\
\de A_i &= \d_i \chi,\\
\de h_{\mu\nu} &= -\d_\mu \xi_\nu - \d_\nu \xi_\mu.
\eal
\eeq
In fact, the quadratic terms in \Eq{gaugedgraveffghostL} are the
most general ones invariant under these symmetries, and armed with the foresight to integrate out $A_0$, we could have used \Eq{gaugedtransformlaws} as the starting point for our analysis.  Indeed, this is exactly what we suggested in Sec.~\ref{sec:bottomup} when we posited the existence of a $a_i$ field.
One advantage of thinking of  \Eq{gaugedgraveffghostL} as coming from gauged ghost condensation is that the weak gauging procedure explains why $g_t$ and $g_s$ are nearly degenerate.

The modified dispersion relation for the scalar excitation can be easily
derived. Parameterizing $\scr{A}_i =\partial_i \sigma$ and choosing
Newtonian gauge, we have
\beq[kinetic]
\scr{L} = \frac{1}{2} \pmatrix{\sigma/g & \Phi^c \cr} 
\pmatrix{ \omega^2 \vec{k}^2- \alpha g^2 \vec{k}^4 & 
-i\epsilon \omega \vec{k}^2 \cr 
i\epsilon \omega \vec{k}^2 &
-\vec{k}^2 (1-\epsilon^2) \cr} 
\pmatrix{\si/g \cr \Phi^c \cr} + \cdots,
\eeq
where
\beq
\epsilon = \frac{M}{ \sqrt{2} \, g \MP},
\eeq
and we have taken the small $g$ limit and dropped the subleading terms
suppressed by $g^2$.%
\footnote{The second term in the $(1, 1)$ entry gives the leading
$\vec{k}^2$ term in the dispersion relation and therefore must be kept.}
Setting the determinant of the kinetic matrix to zero, we
obtain
\beq
\omega^2 = c_s^2\, \vec{k}^2,
\eeq
where
\beq
c_s^2 = \alpha (g^2- g_c^2),
\qquad
g_c = \epsilon g = \frac{M}{\sqrt{2}\MP}.
\eeq
We can see that if $g> g_c$, this is a healthy dispersion relation and there
is no instability even at arbitrarily long wavelengths.
If we also keep the leading contribution at $\scr{O}(\vec{k}^4/M^2)$,
then the dispersion relation is
\beq
\omega^2 = c_s^2\, \vec{k}^2 + \alpha \frac{\vec{k}^4}{M^2},
\eeq
which reproduces the behavior of ghost condensation for $|\vec{k}| \gg g M$.

The dispersion relation of the tensor modes is also modified
because $(\nabla_i {\scr{A}}_j)^2$ contains a term $ \frac 12 \avg{\scr{A}_0}
\d_t h_{ij}$, which contributes to the time derivative of $h_{ij}$.
As a result, the speed of usual
gravitational waves is modified by $\scr{O}(\alpha_2 M^2/\MP^2)$.
If the speed of the gravitational wave is smaller than the speeds of
ordinary particles, there is a strong constraint from cosmic rays
\cite{Moore:2001bv,Elliott:2005va}.

\subsection{Modification of Gravity and Observational Constraints}
Because of the mixing between the Goldstone boson and the metric tensor,
the gravitational potential is modified.  In the static limit,
$\omega \to 0$, the only modification is a simple renormalization of
Newton's constant,
\beq[GNren]
G_N = \frac{G^0_{N}}{1-\epsilon^2} = \frac{1}{8\pi \MP^2(1-\epsilon^2)},
\eeq
as one can see directly from \Eq{kinetic}.
By itself, this does not lead to any measurable effects.

There are non-trivial effects
if the source is moving relative to the ether rest frame, as expected
in realistic situations.
For a source of mass $M_\odot$ moving with velocity $\vec{v}$ relative
to the ether rest frame, the stress-energy tensor is
\beq
T_{00}=\rho= M_\odot \delta^{(3)}(\vec{x}-\vec{v}t),
\eeq
and $T_{0i}$
and $T_{ij}$ are down by powers of $v=|\vec{v}|$ and can be ignored for
$v \ll 1$.   The gravitational potential can be easily obtained through
the $\Phi\Phi$ propagator by inverting the kinetic matrix in \Eq{kinetic},
\beq
\langle \Phi\Phi \rangle  = -\frac{1}{2 \MP^2}\, \frac{1}{\vec{k}^2}
\left( 1- \frac{\alpha g^2 \epsilon^2 \vec{k}^2}{\omega^2 -\alpha g^2
(1-\epsilon^2)\vec{k}^2}\right).
\eeq
The Fourier transform of the source is
\beq
\tilde{\rho}(\omega, \vec{k}) = 2\pi M_\odot \delta(\omega -
\vec{k}\cdot\vec{v}).
\eeq
The Newtonian potential is obtained by performing 
the inverse Fourier transform of the $\langle \Phi\Phi \rangle$ propagator
after substituting $\omega$ by $\vec{k}\cdot\vec{v}$.
Depending on the source velocity in the preferred frame $v$ relative to the 
velocity of the scalar Goldstone mode $c_s \equiv \sqrt{\alpha (g^2-g_c^2)}$,
we get very different results. It is important to distinguish the two different
cases

For $v < c_s$, we can expand
in powers of $v/c_s$ and  obtain the 
angular-dependent gravitational potential
\beq
V(r, \theta) = -\frac{G^0_{N} M_\odot}{r} \left[1 + \frac{\epsilon^2}{1-\epsilon^2} \left(1+ 
\frac{v^2}{2c_s^2}\sin^2 \theta + \scr{O}(v^4/c_s^4)\right) \right],
\eeq
where 
$\cos \theta = \hat{r}\cdot \hat{v}$ is the angle measured from the direction of the source
moving with respect to the ether.   In the parametrized post-Newtonian (PPN) formalism~\cite{Will:1981cz}
this velocity effect with
respect to the preferred frame corresponds to the PPN parameter 
$\alpha_{2}^{\rm PPN}$,
\beq[alphatwoequation]
\alpha_{2}^{\rm PPN}= \frac{\epsilon^2}{(1-\epsilon^2) c_s^2} = 
\frac{M^2}{2\alpha g^4 (1-\epsilon^2)^2 \MP^2}.
\eeq
The observational bound on $\alpha_{2}^{\rm PPN}$ is
\beq[alpha2bound]
\alpha_{2}^{\rm PPN} < 4\times 10^{-7}
\eeq
from the alignment of the solar spin axis and its 
ecliptic~\cite{Nordtvedt:1987,Will:2001mx,Will:1981cz}. 
This provides the strongest bound on
the scale of Lorentz symmetry breaking in this case.\footnote{There
is also a bound from the rate of the cosmic expansion during the 
Big Bang nucleosynthesis due to the different effective Newton's constant for
the cosmological evolution~\cite{Carroll:2004ai}, but it is much weaker.}
For $\alpha g^4$ close to 1 we have $M \lsim 10^{15}$ GeV, and the constraint
is stronger ($ M \lsim 10^{15} \sqrt{\alpha} g^2$ GeV) for smaller $g$.
This bound was also considered in Ref.~\cite{Graesser:2005bg} using the parametrization
of Ref.~\cite{Eling:2003rd} (which corresponds to a decoupling limit of our theory as shown in App.~\ref{app:decoupling}). 
A more systematic derivation of the modification of gravity and discussion
of other PPN parameters can be found in Appendices~\ref{app:Tmunu}~and~\ref{app:ppn}.  In particular, we find that $\alpha_{2}^{\rm PPN}$ is the only modification of General Relativity at post-Newtonian order, showing that gauged ghost condensation yielding a very mild modification of gravity. Our analysis also exhibits the applicable range of the PPN formalism in this theory. {\em It is not enough just for $v$ to be small because the expansion parameter is actually $v/c_s$.} It requires $v< c_s$ and the mixing parameter $\epsilon$ to be small. The above result holds for any theory where Lorentz invariance is broken by the vev of a vector field (such as \Refs{Will:1972,Hellings:1973,Jacobson:2001yj,Eling:2003rd,Jacobson:2004ts,Kostelecky:1989jw,Kostelecky:2003fs,Clayton:1998hv,Clayton:2001vy,Moffat:1992ud,Moffat:2002nm,Gripaios:2004ms,Bertolami:1998dn}).

On the other hand, if $v >c_s$, the PPN expansion breaks down. We expect
that there will be \v{C}erenkov radiation into the Goldstone field. By causality,
modifications of gravity happen only in a cone with an angle
$\theta = \sin^{-1} (c_s/v)$ behind the source.\footnote{This is true for distances larger than $(gM)^{-1}$. For $gM< |\vec{k}|< M$, the $\vec{k}^4$ term in the dispersion relation becomes more important than the $\vec{k}^2$ term, and the scalar velocity is enhanced and becomes $\vec{k}$ dependent, $c_s \sim \sqrt{\alpha} |\vec{k}|/M$. The distance scale where this is relevant is very small
if $M$ is high and $g$ is not too small.}
For $v \gg c_s$, the cone is narrow and
we may not see modifications
of gravity in the solar system where the most stringent bounds come from,
if the cone lies outside the ecliptic plane.
(Of course, there will be some anomalous acceleration if an astrophysical 
object happens to move into the cone of shadow of a gravitational source.)
The most interesting effects in this case probably come from the energy 
loss due to the \v{C}erenkov radiation. Because the coupling of the Goldstone
mode to gravity is proportional to $M$, the \v{C}erenkov radiation due to
gravity scales as $M^2$. The ability to raise the Lorentz symmetry breaking
scale $M$ in gauged ghost condensation makes this effect interesting.
In fact, it provides a significant bound on $M$. In contrast, in ungauged
ghost condensation, the bound $M \lsim 100$ GeV renders this effect totally
irrelevant. However, to discuss the \v{C}erenkov radiation from moving
stars or planets, we needs to first understand the nonlinear effects and where
they become important, so we postpone the discussion of
\v{C}erenkov radiation to Sec.~\ref{sec:Cerenkov} after we discuss 
nonlinear effects.

\section{Nonlinear Effects}
\label{sec:nonlinear}
Up to now we have confined our analysis to the linear order in the fields.
In this section, we show that nonlinear effects can be important for
sufficiently strong gravitational fields, and analyze the dynamics in
the nonlinear regime both analytically and numerically.
We show that the nonlinear dynamics has a simple physical
interpretation as the dynamics of a charged fluid.
We use this picture to give a qualitative understanding of the
nonlinear dynamics.
In particular, we argue that the
caustic singularities found in ungauged ghost condensation
are not present in the gauged case.

\subsection{The Nonlinear Regime}
Recall that
in ungauged ghost condensation, nonlinear
effects are important for all interesting gravitational sources, such
as stars and galaxies~\cite{Arkani-Hamed:2005gu,Krotov:2004if}.
The physical reason for this is that the ghost condensate gravitates
like a fluid that obeys the equivalence principle,
and therefore has a gravitational response time of order
\beq
t_{\rm grav} \sim \frac{r}{\sqrt{\Phi}}.
\eeq
where $r$ is the distance to the gravitating source.
This effect appears only at nonlinear order;
at linear order, the ghost condensate has a highly suppressed
response time due to the $\om^2 = \al \vec{k}^4 / M^4$ dispersion
relation:
\beq
t_{\rm linear} \sim \frac{M r^2}{\sqrt\al}.
\eeq
When $t_{\rm grav} \lsim t_{\rm linear}$, the nonlinear effects
dominate.
These nonlinear effects were studied analytically and
numerically in \Ref{Arkani-Hamed:2005gu}, and it was shown
that regions with $X - 1 < 0$ generally shrink to
small size, resulting in a breakdown of the effective theory.
This is not necessarily a disaster for the theory, since the
energy involved in these singular regions is very small,
but it is unfortunate that this behavior cannot be understood
in the effective theory.

In the case of gauged ghost condensation,
we expect that it also gravitates on a timescale
$t_{\rm grav}$, but the linear timescale is given by
the more conventional dispersion relation $\om^2 = c_s^2 \vec{k}^2$:
\beq
t_{\rm linear} \sim \frac{r}{c_s}.
\eeq
We therefore expect that gravitational effects become important when
$t_{\rm grav} \lsim t_{\rm linear}$, or
\beq[stronggravggc]
\Phi \gsim c_s^2.
\eeq
This means that nonlinear effects become important only for
sufficiently strong gravitational fields.
This will be important in subsequent sections
when we discuss bounds on the gauged ghost condensate.

The result \Eq{stronggravggc} can be derived as follows. 
Expanding $X-1$ to include the leading nonlinear interaction,
\beq
\sfrac{1}{2}(X-1) = \frac{\scr{A}_0}{M} - \Phi -\frac{1}{2M^2} 
\vec{\scr{A}}\,{}^2 +\cdots,
\eeq
the leading terms in the effective Lagrangian are
\beq
\scr{L} = \frac{M^2}{2}\left(\scr{A}_0 - M\Phi -\frac{1}{2M} 
\vec{\scr{A}}\,{}^2\right)^2
- \frac{1}{4g^2} F_{\mu\nu} F^{\mu\nu} - \frac{\alpha}{2}
(\grad \cdot \vec{\scr{A}}\,)^2.
\eeq
Integrating out the massive $\scr{A}_0$ mode,
\beq
\scr{A}_0 = M\Phi + \frac{1}{2M} \vec{\scr{A}}^2,
\eeq
we obtain
\beq
\frac{1}{2g^2} F_{0i}^2 = \frac{1}{2g^2} E_i^2,
\eeq
where we define the ``electric field''
\beq[Edefinition]
\vec{E} = \d_t \vec{\scr{A}}
- M \grad \Phi - \frac{\grad (\vec{\scr{A}}\,{}^2)}{2M}.
\eeq
The leading terms in the nonlinear effective Lagrangian are therefore
\beq[ggcnonlinearlagrangian]
\scr{L} = \frac{1}{2g^2} E_i^2
- \frac{1}{4 g^2} F_{ij}^2 
- \frac{\al}{2} (\grad \cdot \vec\scr{A}\,).
\eeq
The leading nonlinear effects are contained in the
$\grad (\vec{\scr{A}}\,{}^2)$ term in $\vec{E}$.

Nonlinear effects become important when the $\grad (\vec{\scr{A}}\,{}^2)$
term becomes comparable to the time derivative term:
\beq
t_{\rm nonlinear} \sim \frac{M r}{\vec\scr{A}}.
\eeq
The field amplitude $\vec\scr{A}$ induced by a gravitational field
$\Phi$ is given by
\beq
\vec\scr{A} \sim M \sqrt{\Phi},
\eeq
which gives
\beq
t_{\rm nonlinear} \sim \frac{r}{\sqrt\Phi} \sim t_{\rm infall},
\eeq
as anticipated.

\subsection{Fluid Picture}
We now show that the nonlinear dynamics of the effective Lagrangian
\Eq{ggcnonlinearlagrangian} has a natural fluid interpretation,
similarly to the one found for ungauged ghost condensation in
\Ref{Arkani-Hamed:2005gu}.
This will be very useful in understanding the nonlinear dynamics
of the theory.

We begin by noting that in the relativistic formulation,
the field $\scr{A}_\mu$ defines the local preferred rest
frame.
In the effective theory with $\scr{A}_0$ integrated out,
the theory naturally defines a fluid with a local 3-velocity given by
\beq
v_i = -\frac{\scr{A}_i}{M}.
\eeq
The definition of $\vec{E}$ \Eq{Edefinition} can then be written as
\beq[N2fluidpicture]
\frac{D v_i}{D t} = -\d_i \Phi
+ \frac{1}{M} \left( - E_i + F_{ij} v_j \right),
\eeq
where
\beq
\frac{D}{Dt} = \d_t + \vec{v} \cdot \grad
\eeq
is the time derivative along the worldline of a fluid particle
(also called the convective or Lagrangian derivative).
\Eq{N2fluidpicture} has the form of Newton's law for a fluid
particle, with gravitational, electric, and magnetic forces
on the \rhs.

The equations for the electric and magnetic fields arise from
the $\scr{A}_i$ equation of motion of the effective Lagrangian,
and can be written as
\beq[ggcampereeom]
\d_t E_i + \d_j F_{ij} = -v_i \d_j E_j - \al g^2 M \d_i \d_j v_j.
\eeq
This has the form of
Amp\'ere's law with an unconventional current density on the \rhs.
Another difference from conventional electrodynamics is the absence of
Gauss' law.
This is expected, since the Higgs mechanism implies that all 3 components
of $\scr{A}_i$ represent physical degrees of freedom.

As discussed in the previous subsection, for sufficiently strong
gravity we can neglect the $\al g^2$ term in \Eq{ggcampereeom},
at least as long as spatial gradients are not too large.
In this case, there is a simple solution with
$E_i = 0$, $F_{ij} = 0$ in which the fluid particles follow gravitational
geodesics.
This is the relevant solution in the case where the fluid
is initially ``at rest'' in the presence of a gravitating source.
Just as in the ungauged case, the fluid particles will ``fall in''
toward the source, giving a very direct and physical picture of
why the gravitational
time scale is relevant for the nonlinear dynamics.
The subsequent evolution of the fluid will in general give rise
to caustic singularities.
Near these caustic singularities, the higher-derivative
$\al g^2$ is important, and may resolve the singularity.
In the fluid picture, the ``electric'' and ``magnetic'' forces
are becoming important near the would-be caustic, and may
cause the incoming fluid particles to ``bounce.''
This is the question we address next.

\subsection{Would-be Caustics}
\label{subsec:Caustics}
To address the question of caustic singularities in the nonlinear
dynamics, we restrict attention to situations with
spherical, cylindrical, or planar symmetry.
We can treat these simultaneously by using the spatial metric
\beq
ds^2 = dr^2 + r^2 d\Om_s^2 + \sum_{p = 1}^{2-s} dx_p^2,
\eeq
where
\beq
d\Om_s^2 = 
\begin{cases}
0 & $s=0$ (planar symmetry) \cr
d\th^2 & $s=1$ (cylindrical symmetry) \cr
d\theta^2+\sin^2\theta d\varphi^2 &
$s=2$ (spherical symmetry) \cr
\end{cases}
\eeq
Since only $\scr{A}_r$ is nonzero, we have $F_{ij} = 0$,
and ``Newton's law'' simplifies to
\beq[newtonlawspherical]
\frac{D v_r}{Dt} = -\d_r \Phi - \frac 1M E_r,
\eeq
where $D / Dt = \d_0 + v_r \d_r$.
``Amp\'ere's law'' \Eq{ggcampereeom} can then be written in the form
\beq[ampereeomspherical]
\frac{D}{Dt} (r^s E_r) = -r^s j_r,
\eeq
where 
\beq
j_r = \al g^2 M \d_r \left[ \frac{1}{r^s} \d_r (r^s v_r) \right].
\eeq
\Eq{ampereeomspherical} says that the comoving
``flux'' $r^s E_r$ is changing according to the ``current density'' $j_r$.
In particular, when $j_r$ is negligible, the ``flux'' is conserved,
a fact that is useful in interpreting the numerical results below.

We now consider an initial condition that can lead to a
caustic singularity.
That is, we assume that initially $v_r < 0$ near $r = 0$,
corresponding to a situation where the fluid particles
are heading toward the origin.
More precisely, we assume that near $r = 0$, we have
\beq
v_r = -c_1 r + c_2 r^2 + \scr{O}(r^3),
\eeq
with $c_1 > 0$.
Because we expect the linear growth of $v_r$ away from $r = 0$
to degrease with $r$, we also assume $c_2 > 0$.
In this case, we find that near $r = 0$
\beq
j_r = (s+2) \al g^2 M c_2 > 0.
\eeq
Therefore, \Eq{ampereeomspherical} implies that
the ``electric field'' $E_r$ tends to \emph{decrease}
with time.
In particular, if $E_r = 0$ initially (\eg\ if the fluid is
``at rest'') then later $E_r < 0$.

The sign of $E_r$ is crucial for the fate of the would-be caustic.
For $E_r < 0$, the ``electric'' force in \Eq{newtonlawspherical}
is repulsive and can cause the incoming particles to bounce.
To see the condition for this, we rewrite \Eq{newtonlawspherical}
as a function of the comoving radial coordinate $r$ as
\beq
\frac{D}{Dr} \left( \sfrac 12 v^2 \right) = -\frac 1M E_r,
\eeq
where we have neglected the gravitational ``force.''
We therefore have
\beq
\De \left( \sfrac 12 v^2 \right) = -\frac 1M \int_{r_i}^{r_f} dr\, E_r.
\eeq
This can be interpreted as the work-energy theorem for fluid particles.
If $E_r < 0$ and the ``work'' integral on the \rhs\ diverges as $r_f \to 0$,
then the particles will bounce.
This occurs for $s = 1, 2$ (cylindrical or spherical symmetry),
since $E_r \sim r^{-s}$ neglecting the
``current'' contribution in \Eq{ampereeomspherical}.
Physically, it requires an infinite amount of energy to compress
a charged sphere or cylinder to zero size.
For $s = 0$ (planar symmetry), the ``work'' integral  is finite,
corresponding to the fact that a finite amount of energy is sufficient
to compress a plane of charge to zero size.
Indeed numerical simulations (discussed below) confirm that
caustic singularities do form for planar symmetry, but not for
cylindrical or spherical symmetry.
However, we expect that departures from perfect planar symmetry
will be important near the would-be caustic.
Specifically, small fluctuations of the plane symmetric
collapse will grow, and we expect the layer to fragment before
a planar caustics occurs.
(See App.~\ref{app:fragmentation} for a 
perturbative analysis supporting this picture.)
After the fragmentation,
lower-dimensional caustics do not occur by the argument above.%

The fact that the ``current density'' $j_r$
generates a negative $E_r$ depends
crucially on the non-vanishing second derivative of $\scr{A}_r$ at
$r = 0$.
In fact, the nonlinear equations have exact solutions
\beq[exactscalingsoln]
A_r = \frac{\scr{C}_s r}{t_c - t},
\quad
E_r = \frac{\scr{C}_s (1 - \scr{C}_s) r}{(t_c - t)^2},
\eeq
where $\scr{C}_s = 1$ or $2/(s+1)$ and
$t_c$ is an arbitrary constant.
For these solutions $j_r = 0$, and nothing prevents the solutions
from collapsing at $t = t_c$.
These solutions are of course unrealistic because of the boundary
conditions at $r \to \infty$, but we might worry that they are
approximately valid near the origin.
However, we expect that the scaling solutions \Eq{exactscalingsoln}
are dynamically avoided.
This is because the second derivative of $A_r$ must be negative
away from $r = 0$, and this propagates to the origin, making
$j_r > 0$ and preventing collapse.

\subsection{Numerical Results}
\begin{figure}
\centering\leavevmode\epsfysize=8cm
\epsfbox{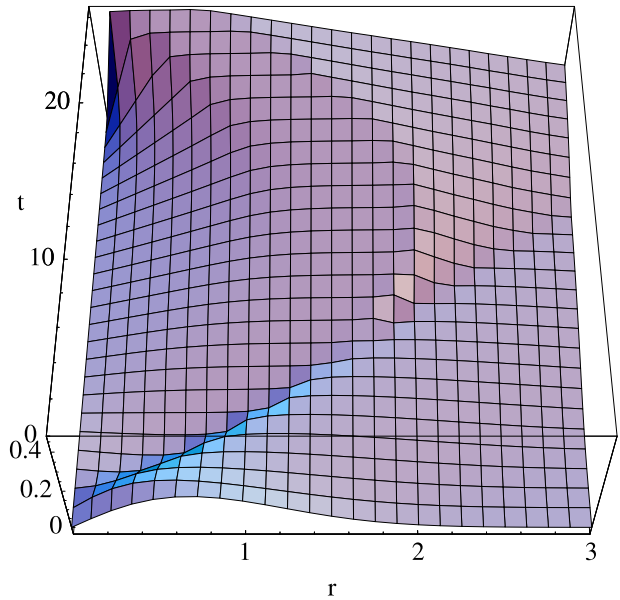}
\caption{\label{fig:plotAr_plane}%
A result of numerical simulation for planar symmetry ($s=0$). The
variable $A_r$ is plotted as a function of $r$ and $t$.  After an initial bounce, a caustic singularity forms at late times.  Note that in a realistic situation, we do not expect a system to exhibit planar symmetry.}
\end{figure}

We now turn to numerical investigations of the nonlinear equations.
We begin with the plane symmetric case ($s=0$) with no external
gravitational potential ($\Phi=0$).
For simplicity we assume the 
symmetry under reflection $r\to -r$. In this case $A_r$ and $E_r$ are
odd functions of $r$.
As the initial condition at $t = 0$ we set
%
\begin{equation}
 A_r|_{t=0} = 2\Omega r e^{-r^2}, \quad E_r|_{t=0} = 0,
\end{equation}
where $\Omega$ is a positive constant. 
By performing numerical simulations, we find that
for some values of $\alpha g^2$ and $\Omega$ the would-be caustic
bounces, but much later it recollapses and forms a caustic that
exits the realm of validity of the effective field theory before
it bounces.
This is shown in Fig.~\ref{fig:plotAr_plane}.
For other values of $\alpha g^2$ and $\Omega$, even the first caustics
at $t\sim 1/\Omega$ does not bounce until the system exits the regime of
validity of the effective field theory.
This confirms the conclusion of the analytical argument above.
As discussed there, we do not expect planar caustics to form in
the realistic case.

\begin{figure}
\centering\leavevmode\epsfysize=8cm 
\epsfbox{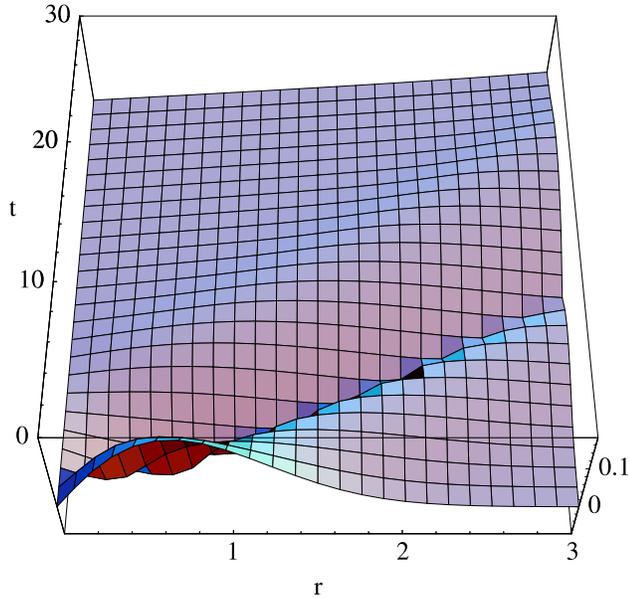}
\caption{\label{fig:plotAr_cyl}
 A result of numerical simulation in cylindrial symmetry ($s=1$). The
 variable $A_r$ is plotted as a function of $r$ and $t$.   After an initial bounce, no caustics form at late times.
 }
\end{figure}

\begin{figure}
 \centering\leavevmode\epsfysize=8cm 
 \epsfbox{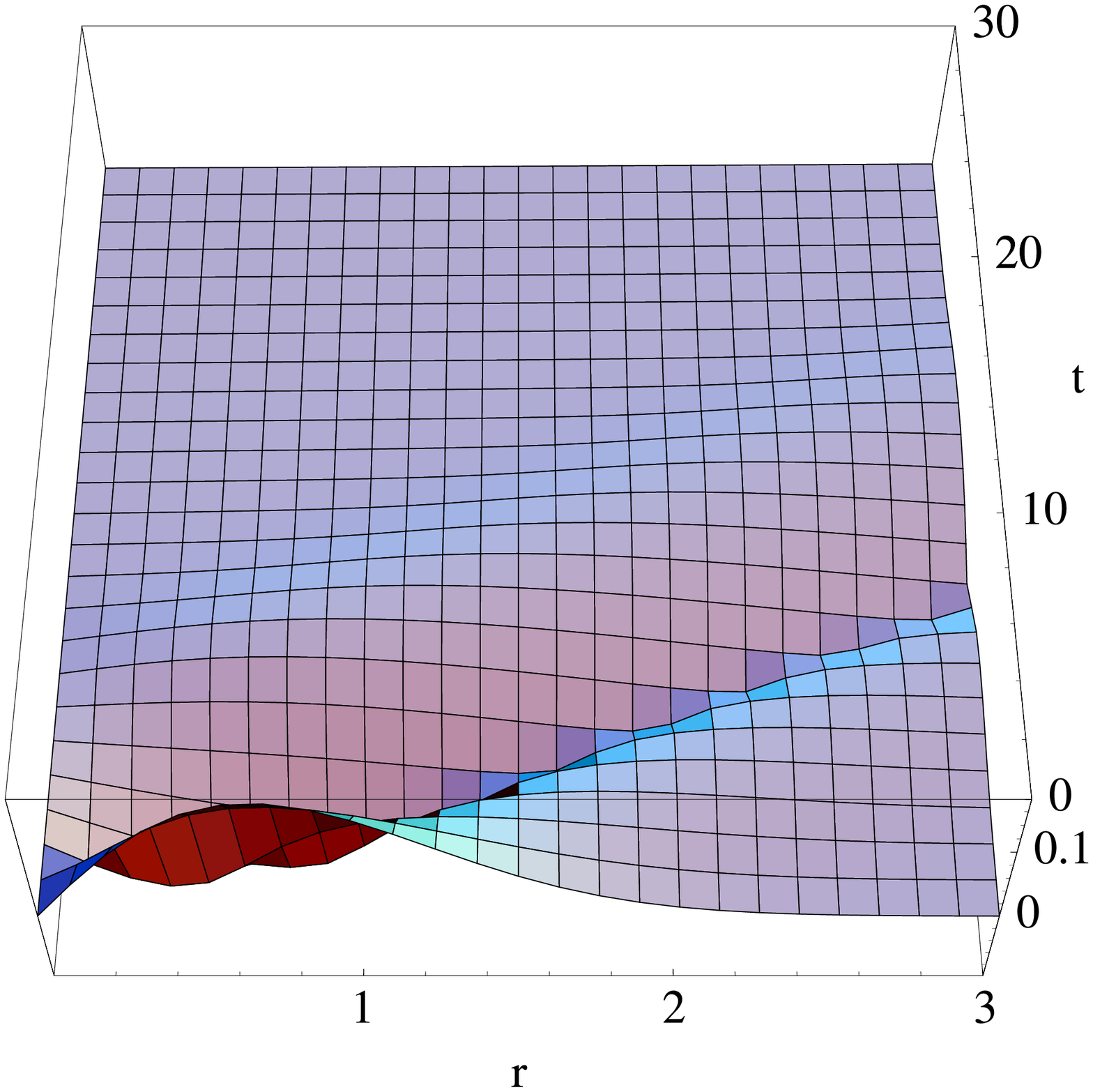}
 \caption{\label{fig:plotAr_sph}
 A result of numerical simulation in spherical symmetry ($s=2$). The
 variable $A_r$ is plotted as a function of $r$ and $t$.  Like the case of cylindrical symmetry, there are no late time caustics.
 }
\end{figure}

We now turn to systems with cylindrical and spherical symmetry.
Figs.~\ref{fig:plotAr_cyl} and \ref{fig:plotAr_sph} show results of
numerical simulation in cylindrical and spherical cases with $\Phi=0$,
respectively.
It is evident that the excitation near the origin diffuses and the system
asymptotically approaches the trivial solution $A_r, E_r = 0$.
These results confirm our expectation that caustics with cylindrical
or spherical symmetry do not occur.

\begin{figure}
 \centering\leavevmode\epsfysize=8cm 
 \epsfbox{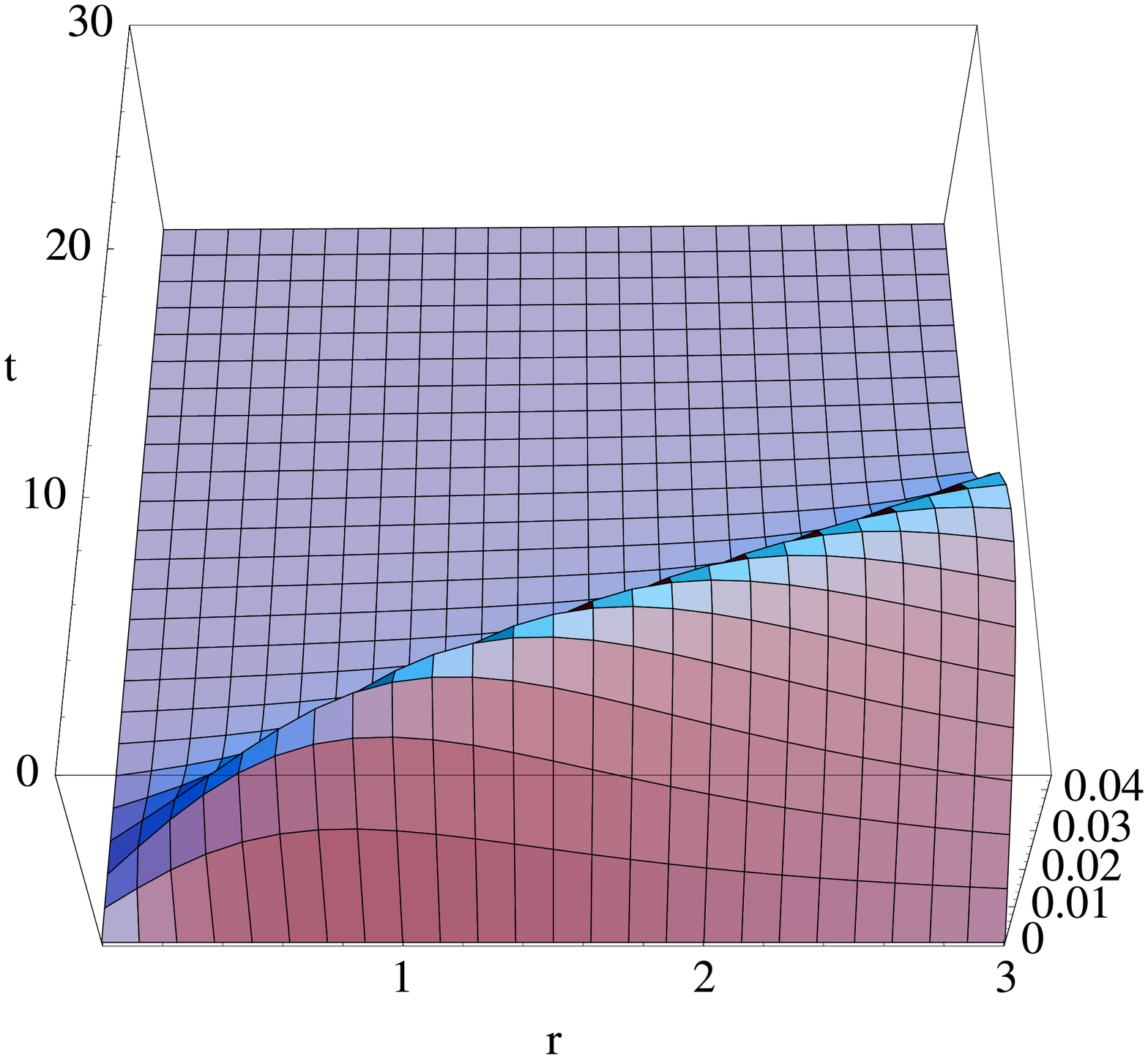}
 \caption{\label{fig:plotAr_grav}
 A result of numerical simulation in spherical symmetry ($s=2$) with an
 external gravitational potential. The variable $A_r$ is plotted as a
 function of $r$ and $t$.  Again, we observe the absence of late time caustics.
 }
\end{figure}

\begin{figure}
 \centering\leavevmode\epsfysize=8cm 
 \epsfbox{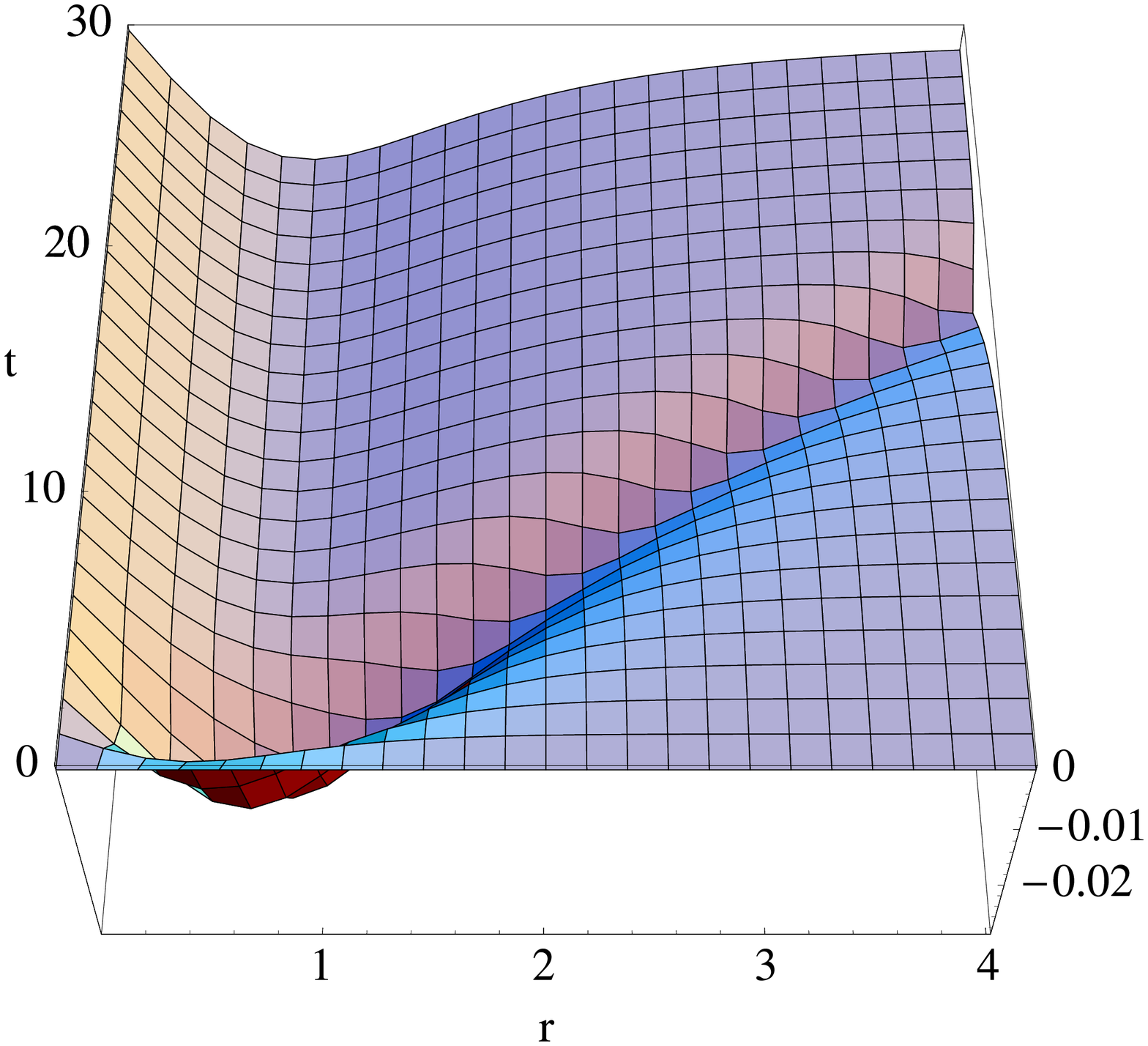}
 \caption{\label{fig:plotEr_grav}
 A result of numerical simulation in spherical symmetry ($s=2$) with an
 external gravitational potential. The variable $E_r$ is plotted as a
 function of $r$ and $t$.  At late times, the ``electric field'' cancels the gravitational force, halting the accretion of the gauged ghost condensate and protecting the system from caustics.
 }
\end{figure}

So far, we have been working in the absence of gravity.
Let us now consider effects of an external gravitational potential.
We still neglect gravity generated by excitations 
of the gauged ghost condensate.
For simplicity we assume spherical symmetry $(s=2)$.
A typical numerical result is shown
in Figs.~\ref{fig:plotAr_grav} and \ref{fig:plotEr_grav}.
We see that in the final stationary configuration,
the repulsive ``electric force'' $E_r$ cancels
the attractive gravitational force $\partial_r\Phi$
(\ie\ $A_r \to 0$, $E_r \to -\partial_r\Phi$).

\section{Black hole accretion}
\label{sec:BH}

In this section we analyze perturbations of Schwarzschild geometry and 
estimate the rate of mass increase of the black hole due to accretion of
the gauged ghost condensate, extending the analysis of
\Ref{Mukohyama:2005rw} for the ungauged ghost condensate.

Before starting the discussion of black holes, let us first consider 
a spherically symmetric, static star surrounded by a gauged ghost condensate. In the previous section we have seen that the gauged
ghost condensate with spherical symmetry quickly approaches a regular,
stationary configuration with or without an external gravitational
potential. In the fluid picture, each fiducial fluid particle follows
``Newton's law'' and approaches the center within the dynamical time set
by the external gravitational potential. Near the center, the ``electric
field'' $E_r$ builds up and the infalling flow of fluid particles
bounces. Part of outgoing flow produced by the bounce may come back to
the center at late time but bounces again. After several (if not one)
bounces, the system settles down to a regular, stationary
configuration. Inclusion of gravitational backreaction just induces the
``renormalization'' of Newton's constant (see \Eq{GNren} for the
renormalization in the linear theory) and does not change this
qualitative behavior towards the stationary configuration since the
stress-energy tensor of the final stationary configuration is
time-independent and does not have any time-space components. Note that
for the ghost fluid around a star to settle down to the stationary
configuration, the existence of a regular center is crucial: bounces
happen because of the boundary condition at the center.

The ghost condensate surrounding a black hole behaves very differently
from that around a star. This is because the boundary condition at the
horizon of a black hole is completely different from that at the center
of a star. Fiducial fluid particles just go through the horizon and 
continue to infall. The flow of ghost fluid should still be regular from
the viewpoint of infalling comoving observers: shell-crossing type
singularities are avoided by bounces due to temporal build-up of the
local ``electric field''. Thus, if we neglect the gravitational
backreaction, the ghost condensate surrounding a black hole should
approach a congruence of free-falling geodesics in the fluid picture
and this should happen roughly within the dynamical (or Kepler)
time. The stress energy tensor vanishes if $\alpha=0$ and is suppressed
by the factor $\alpha M^2/M_{Pl}^2$. Hence, the gravitational
backreaction does not change this qualitative behavior happening within
the dynamical time. Thus, the congruence of free-falling geodesics is a
very good approximation to the full solution including gravitational 
backreaction. However, if we are interested in physics of much longer
time scales, the gravitational backreaction may build up to appreciable
amount. Therefore, we would like to estimate the mass
increase of a black hole due to accretion of the ghost
condensate. In the following, we first find the approximate
solution representing the congruence of free-falling geodesics and then
estimate the gravitational backreaction by perturbative expansion
with respect to $\alpha M^2/M_{Pl}^2$.

In order to investigate the behavior of the gauged ghost condensate in a
black hole background, we need to choose the time variable
carefully. For example, if we use the background Killing time as a time
variable then there would appear infinite blue-shift on a black hole
horizon. With the infinite blue-shift, the local physical energy scale
near the horizon would easily go beyond the scale $M$. In particular, we
would not be allowed to integrate out the field $A_0$, which has an
order-1 overlap with the massive mode in Eq.~(\ref{eqn:heavy-mode}). For this
reason we should use a different time coordinate for which there is no
large blue-shift. In the following analysis we shall use an infalling
Gaussian normal coordinate system, in which there is no blue-shift since
the time-time component of the metric is $-1$ everywhere by
definition. In this coordinate we can safely use the effective theory
with $A_0$ integrated out. As explained in the previous paragraph, the
gauged ghost condensate surrounding a black hole quickly settles down to
a congruence of infalling geodesics in the fluid picture. In other
words, the solution in the zeroth order in $\alpha M^2/M_{Pl}^2$ is
given by $\phi=M^2\tau$, where $\tau$ is the time coordinate in the
infalling Gaussian normal coordinate system.

The unperturbed Schwarzschild metric in the infalling Gaussian normal
coordinate called the Lemaitre reference frame~\cite{Lemaitre:1933qe} is 
%
\begin{equation}
 ds^2 = -d\tau^2 + \frac{(2m_0)^3}{r(\tau,x)}dx^2
  + r^2(\tau,x)d\Omega^2, 
\end{equation}
where
%
\begin{equation}
 r(\tau,x) = 
  2m_0
  \left[\frac{3}{2}\left(x-\frac{\tau}{2m_0}\right)\right]^{2/3}.
  \label{eqn:def-r}
\end{equation}
There is nothing bad on the future (black hole) horizon and the
coordinate system covers everywhere in the region $v>-\infty$, where $v$
is the ingoing Eddington-Finkelstein null coordinate~\cite{MTW}. The
metric becomes ill only on the curvature (physical) singularity at
$\tau=2m_0 x$. 
We consider spherically-symmetric, time-dependent perturbations of the
Schwarzschild geometry. We shall still use an infalling Gaussian normal
coordinate system:
%
\begin{equation}
 ds^2 = -d\tau^2 + \frac{(2m_0)^3e^{2\delta_1(\tau,x)}}{r(\tau,x)}dx^2
  + e^{2\delta_2(\tau,x)}r^2(\tau,x)d\Omega^2, 
\end{equation}
where $r(\tau,x)$ is given by (\ref{eqn:def-r}). We consider
$\delta_1$ and $\delta_2$ as perturbations. Up to the linearized level,
the Misner-Sharp energy~\cite{Misner:1964je} is expanded as
%
\begin{equation}
 M_{MS} = \frac{e^{\delta_2}r M_{Pl}^2}{2}
  \left[ 1 - 
   \partial^{\mu}(e^{\delta_2}r)
   \partial_{\mu}(e^{\delta_2}r)\right]
  = M_{Pl}^2(m_0 + m_1),
\end{equation}
where
%
\begin{equation}
 \frac{m_1}{r} = \delta_1 - 2m_0\left(\frac{r}{2m_0}\right)^{1/2}
  \partial_{\tau}\delta_2
  - \left(\frac{r}{2m_0}\right)^{3/2}\partial_{x}\delta_2
  - \left(1-\frac{3m_0}{r}\right)\delta_2.
\end{equation}
Note that the Misner-Sharp energy agrees with the ADM energy at large
distances and, thus, measures the strength of gravity. In the following
we shall eliminate $\delta_1$ from all equations by using this
expression and consider $m_1$ and $\delta_2$ as dynamical variables. As
for the gauged ghost condensate, we introduce linear perturbation
$A(\tau,x)$ as $A_idx^i=A(\tau,x)dx$. Einstein's equation gives a
set of coupled partial differential equations for the three variables
$m_1$, $\delta_2$ and $A$. Fortunately, it is possible to decouple
$\delta_2$ from equations for $m_1$ and $A$ by taking the following
linear combinations of the linearized Einstein tensor: 
%
\begin{eqnarray}
 (2m_0)^2\left(\frac{r}{2m_0}\right)^{3/2}
 \left[G_{\perp\perp}+\frac{G_{\perp x}}{2m_0}\right]
 & = & \frac{\partial_{x}m_1}{m_0}, \nonumber\\
 (2m_0)^3\left(\frac{r}{2m_0}\right)^{1/2}
  \left[ G^{xx}+\frac{r^2}{(2m_0)^5}G_{\perp x} \right]
  & = & \frac{\partial_{\tau}m_1}{m_0}.
\end{eqnarray}

By applying the formula in Appendix~\ref{app:Tmunu} to 
%
\begin{eqnarray}
 N & = & 1, \quad \beta = 0, \nonumber\\
 q_{ij}dx^idx^j & = & 
  \frac{(2m_0)^3e^{2\delta_1(\tau,x)}}{r(\tau,x)}dx^2
  + e^{2\delta_2(\tau,x)}r^2(\tau,x)d\Omega^2, \nonumber\\
 A_idx^i & = & A(\tau,x)dx, 
\end{eqnarray}
we can obtain the stress energy tensor of the gauged ghost condensate.

In this setup, we expect that accretion of condensate to a black hole is
due to the $\alpha$ term. Hence, we perform a perturbation analysis with respect to
$\alpha M^2/M_{Pl}^2$, considering $m_1$, $\delta_2$ and $A$ as
quantities of the first order in the perturbation. At zeroth order in $\alpha M^2/M_{Pl}^2$, all
components of the stress energy tensor vanish and the Einstein equation
is automatically satisfied. This means that there is no mass increase
when $\alpha M^2/M_{Pl}^2=0$. 

At first order in $\alpha M^2/M_{Pl}^2$,
the non-vanishing components of the stress energy tensor are
%
\begin{eqnarray}
 T_{\perp\perp} & = & -\frac{M^2}{(2m_0)^2}
  \left[ \frac{1}{g^2}\frac{r}{2m_0}\partial_{\tau}\partial_x A
  + \frac{5}{2g^2}\left(\frac{2m_0}{r}\right)^{1/2}\partial_{\tau}A
  +\frac{9\alpha}{8}\left(\frac{2m_0}{r}\right)^3\right], \nonumber\\
 T_{\perp x} & = & \frac{9\alpha M^2}{8m_0}
  \left(\frac{2m_0}{r}\right)^3, \nonumber\\
 T^{xx} & = & -\frac{9\alpha M^2}{8 (2m_0)^4}
  \left(\frac{2m_0}{r}\right)^2, 
  \nonumber\\
 T^{\theta\theta} & = & -\frac{9\alpha M^2}{8(2m_0)^4}
  \left(\frac{2m_0}{r}\right)^5. 
\end{eqnarray}
Hence, we obtain 
%
\begin{eqnarray}
 T_{\perp\perp}+\frac{T_{\perp x}}{2m_0}
 & = & -\frac{M^2}{(2m_0)^2}
  \left[ \frac{1}{g^2}\frac{r}{2m_0}\partial_{\tau}\partial_x A
+ \frac{5}{2g^2}\left(\frac{2m_0}{r}\right)^{1/2}\partial_{\tau}A
  -\frac{9\alpha}{8}\left(\frac{2m_0}{r}\right)^3\right], \nonumber\\
 T^{xx}+\frac{r^2}{(2m_0)^5}T_{\perp x}
  & = & - \frac{9\alpha M^2}{8(2m_0)^4}\left(\frac{2m_0}{r}\right)^2
  \left( 1 - \frac{r}{m_0}\right).
\end{eqnarray}
Thus, Einstein equation implies that
%
\begin{equation}
\frac{1}{g^2}\frac{r}{2m_0}\partial_{\tau}\partial_x A
  + \frac{5}{2g^2}\left(\frac{2m_0}{r}\right)^{1/2}\partial_{\tau}A
  = \frac{9\alpha}{8}\left(\frac{2m_0}{r}\right)^3
  - \frac{M_{Pl}^2}{M^2}\left(\frac{2m_0}{r}\right)^{3/2}
  \frac{\partial_{x}m_1}{m_0},
\end{equation}
and
\begin{equation}
 \partial_{\tau}m_1 = 
  - \frac{9\alpha M^2}{16 M_{Pl}^2}\left(\frac{2m_0}{r}\right)^{3/2}
  \left( 1 - \frac{r}{m_0}\right).
\end{equation}
The first equation should be considered as an equation for $A$ for a
given $m_1$. The second equation does not include $A$ and, thus, can be
easily solved w.r.t. $m_1$. The general solution for $m_1$ is
%
\begin{equation}
 \frac{m_1}{m_0} = \frac{9\alpha M^2}{4M_{Pl}^2}
  \left[ -\frac{r}{2m_0} 
   + \frac{1}{2}\ln\left(\frac{r}{2m_0}\right) + \bar{C}(x)
  \right],
\end{equation}
where $\bar{C}(x)$ is an arbitrary function. Note that $\tau$
parameterizes each geodesic as the proper time measured by a comoving
observer and $x$ parameterizes the congruence of geodesics. Thus, the 
function $\bar{C}(x)$ represents difference between evolution along
different infalling geodesics and thus the initial condition on an
initial spacelike hypersurface.

This result is equivalent to the corresponding expression (22) in
\cite{Mukohyama:2005rw} for the ungauged ghost condensate. To see the  
equivalence, notice that $x$ here corresponds to $x_+$ in
\cite{Mukohyama:2005rw}. Thus, the accretion rate up to first order in
$\alpha M^2/M_{Pl}^2$ is the same as that in the ungauged case. In
particular, as shown in \cite{Mukohyama:2005rw}, the leading late-time
behavior is 
%
\begin{equation}
 \frac{m_1}{m_0} \sim \frac{9\alpha
  M^2}{4M_{Pl}^2}\left(\frac{3v}{4m_0}\right)^{2/3}, 
\end{equation}
where $v$ is the advanced null time coordinate normalized at
infinity (the ingoing Eddington-Finkelstein null coordinate). To obtain
this formula, the asymptotic behavior of the function $\bar{C}(x)$
for large $x$ has been determined by the assumption that the initial
value of $m_1$ on an initial spacelike hypersurface does not diverge at
large $r$, and the limit $v\gg r$ has been taken. This formula is valid
in the regime where $m_1$ is sufficiently small compared to $m_0$. In
order to see the behavior beyond this regime, let us ``renormalize'' the
parameter $m_0$. For this purpose we note that 
%
\begin{equation}
 \frac{1}{M_{Pl}^{4/3}}\frac{d(M_{BH}^{2/3})}{d(v^{2/3})} \sim
  2\left(\frac{3}{4}\right)^{5/3}\frac{\alpha M^2}{M_{Pl}^2}
\end{equation}
for the black hole mass $M_{BH}=(m_0+m_1)M_{Pl}^2$. This gives the
solution 
%
\begin{equation}
 M_{BH}^{2/3} \sim M_0^{2/3} 
  + 2\left(\frac{3}{4}\right)^{5/3}\frac{\alpha M^2}{M_{Pl}^2}(M_{Pl}^2v)^{2/3}
  \label{eqn:MBHincrease}
\end{equation}
for the mass $M_{BH}$ surrounded by the sphere at the radius $r$ at the
time $v$, where $M_0$ is the initial value of $M_{BH}$ at $v=0$. This
formula is valid if the condition $v\gg r$ is satisfied. We shall check
this condition when we apply this formula to a stellar-mass black hole
below.

Convincing evidences for stellar-mass black holes are provided by
X-ray binaries, such as Cygnus X1. Some candidates for the stellar-mass
black holes are listed in Table 1 of \cite{Orosz:2002uh}. On the other
hand, ages of those stellar-mass BHs are less certain. For example, XTE
J1118+480 is thought to be a black hole with mass 
$M_{BH} \sim 7M_{\odot}$ and binary separation $r\sim 3R_{\odot}$, but
its age is estimated to be either $\sim 240$~Myr or 
$\sim 7$~Gyr~\cite{Mirabel:2001ay}, depending on whether it was kicked by
a supernova explosion or was  ejected from a globular cluster. (GRO
J1655-40 has $M_{BH} \sim 5M_{\odot}$ and its age is estimated to be 
$\sim 0.7$~Myr~\cite{Mirabel:2002uc}. This would give a weaker bound.)

The observation of XTE J1118+480 implies that the mass measured at
$r\sim 3R_{\odot}$ is $M_{BH}\sim 7M_{\odot}$ at $t\sim 240$~Myr or 
$\sim 7$~Gyr, while the theory of stellar evolution 
says that $M_0$ must be larger than $\sim 3M_{\odot}$. Note that the
estimate of $t$ should not change significantly even if $M_{BH}$
actually evolved from $\sim 3M_{\odot}$ to $\sim 7M_{\odot}$ during its
journey. We can apply the formula (\ref{eqn:MBHincrease}) to this system
since $M_{Pl}^2r/M_{BH}\sim 10^5$ and $M_{Pl}^2t/M_{BH}\sim 10^{20}$
and thus the condition $v\sim t\gg r$ is satisfied. Therefore, we
obtain 
%
\begin{equation}
 (7M_{\odot})^{2/3} - (3M_{\odot})^{2/3} \gsim 
  \frac{\alpha M^2}{M_{Pl}^2} (M_{Pl}^2\times 240\mbox{ Myr})^{2/3}. 
\end{equation}
This gives the bound~\footnote{
Because of the motion of the binary system, the relative rotation of the
binary components and the non-zero spin of the black hole, the spherical
symmetry assumed in the derivation of the formula
(\ref{eqn:MBHincrease}) is violated. It is expected that these factors
will increase the accretion rate. Thus, while (\ref{eqn:bh-bound}) is
valid as an upper bound on $\sqrt{\alpha}M$, it is worthwhile seeking
more stringent bound by taking them into account. 
}  
on $M$ as
%
\begin{equation}
 \sqrt{\alpha}M \lsim 10^{12} \GeV. 
  \label{eqn:bh-bound}
\end{equation}

\section{\v{C}erenkov Radiation from Gravitating Sources}
\label{sec:Cerenkov}

For a gravitational source moving faster than the sound velocity of the ether
($v > c_s$)
ether \v{C}erenkov radiation will be emitted.
As we will see, the energy loss per unit time is at most of order $M^2 v^3$,
where $M$ is the scale of the gauge ghost condensate,
independently of the size of the source.
In the ungauged ghost condensate, the same result holds \cite{Arkani-Hamed:2004ar},
but in the gauged case $M$ can be much larger, which give rise to interesting
phenomena and constraints from this effect.

Consider a non-relativistic source
\begin{equation}
\label{eq:L_S}
{\cal L_S}= -\frac{1}{\sqrt{2} M_{\rm Pl}} \rho\, \Phi^c .
\end{equation}
As shown in App.~\ref{sec:simpleway}, the energy loss rate can be calculated as 
\begin{equation}
\label{eq:energyloss}
\dot{E}_{\rm src} = \int d^3 x \frac{1}{\sqrt{2}M_{\rm Pl}} \rho(x, t)
\dot{\Phi}^c(x,t) .
\end{equation}
At the linearized level,
\begin{equation}
\label{eq:response}
\tilde{\Phi}^c(\omega, k) = \left[ -\frac{1}{k^2} + \frac{\alpha M^2}{2 M_{\rm Pl}^2}\,\frac{1}{\omega^2- c_s^2 k^2} \right] \frac{\tilde{\rho}(\omega, k)}{\sqrt{2}M_{\rm Pl}} ,
\end{equation}
where we have ignored the $k^4$ piece in the denominator of the second term
in the bracket because it is never  important for the large astrophysical bodies we consider here.

For a star of mass $M_\odot$ moving on a straight line with velocity $\vec{v}$:
\begin{equation}
\label{eq:source}
\tilde{\rho}(\omega, k) = 2\pi M_\odot \delta (\omega - \vec{k}\cdot \vec{v}) f(k),
\end{equation}
where $f(k)$ is the dimensionless form factor given by the Fourier transform of the mass density distribution. For example, a uniform-density sphere of radius $R_\odot$ gives
\begin{equation}
f(k) = \frac{3(\sin kR_\odot -kR_\odot \cos kR_\odot)}{(kR_\odot)^3}.
\end{equation}
Substituting \Eq{source} into \Eqs{energyloss} and \eq{response}, we have 
\begin{equation}
\dot{E}_{\rm src} = \frac{M_\odot^2}{2M_{\rm Pl}^2} \int \frac{d^3 k}{(2\pi)^3}
(-i \vec{k}\cdot \vec{v}) \left[-\frac{1}{k^2} + \frac{\alpha M^2}{2 M_{\rm Pl}^2}\,\frac{1}{(\vec{k}\cdot \vec{v}+ i \varepsilon)^2- c_s^2 k^2} \right]
|f(k)|^2,
\end{equation}
where we have used the $i\epsilon$ prescription for the retarded Green's function. Note that the integral is only nonzero due to the poles of the second term, {\it i.e.}\ when the propagator is on shell. This occurs only for $v> c_s$, and we obtain
\begin{equation}
\label{eq:k_integral}
\dot{E}_{\rm src} = -\frac{\alpha M_\odot^2 M^2}{16\pi M_{\rm Pl}^4 v} \int_0^\infty
k\,dk |f(k)^2|,
\end{equation}
for $v \gg c_s$. For a source of size $R_\odot$, we expect
\begin{equation}
\int_0^\infty k\,dk |f(k)^2| \sim \frac{1}{R_\odot^2}.
\end{equation}

However, as we discussed in the previous section, nonlinear effects
can be important near a gravitational source, and may cut off the integral
\Eq{k_integral} at radius larger than $R_\odot$.
Using the estimates of the previous section, we see that nonlinear effects
become important within a distance
\beq
r \lsim R_{\rm nonlinear} \sim \frac{M_\odot}{\MP^2 c_s^2}.
\eeq
In the ungauged ghost condensate, the parametric formula for the rate of
energy loss in the linear regime continues to be valid in the
nonlinear regime \cite{Arkani-Hamed:2005gu},
and we find the same result here for essentially the same reason.
Consider a fluid particle moving past a source with speed $v$ and
impact parameter $r$.
In the impulse approximation, a fluid particle gets a kick perpendicular
to its initial velocity due to gravity
\beq
\De v_\perp \sim F \De t \sim \frac{M_0}{\MP^2 v r}.
\eeq
Here we have neglected the ``electric'' force on the particle.
This can be estimated using
\beq
\frac 1M E_\perp
\sim \frac{\al g^2 \De v_\perp}{r^2} \De t
\sim \frac{\al g^2 M_0}{\MP^2 v^2 r^2}
\sim \frac{c_s^2}{v^2} \grad \Phi.
\eeq
We see that neglecting the ``electric force'' is justified for $v > c_s$.
The impulse approximation is valid as long as the perpendicular
distance travelled by the fluid particle is less than $r$, which gives
\beq
r \lsim R_{\rm drag} \sim \frac{M_0}{\MP^2 v^2}.
\eeq
Note that $R_{\rm drag} < R_{\rm nonlinear}$ for $v > c_s$.
For $r \lsim R_{\rm drag}$, we expect the fluid to be ``dragged'' with the
source, and no \v{C}erenkov radiation will be emitted from the dragged
region.

The energy loss induces an anomalous acceleration in the direction of the ether wind,
\begin{equation}
a_{\rm anom} = \frac{\dot{E}_{\rm src}}{M_\odot v}
\simeq \frac{ \alpha \kappa M_\odot M^2}{16\pi M_{\rm Pl}^4 v^2 R_\odot^2} .
\end{equation}
Because it is proportional to $M_\odot/R_\odot^2$, the Sun gets the largest anomalous
acceleration in the solar system. These accelerations will modify the orbits of the planets. In particular, they will cause misalignments of the orbital planes if the anomalous accelerations are out of the ecliptic plane. For comparison, the typical relative accelerations between the planets and the Sun are $10^{-34} - 10^{-38}$ GeV, the Viking ranging data constrains any anomalous radial acceleration acting on Earth or Mars to be smaller than $\sim 10^{-44}$ GeV while the Pioneer anomaly corresponds to $a_{\rm Pioneer} \sim 10^{-42}$ GeV~\cite{Anderson:1998jd}. If we require the anomalous accelerations to be smaller than $\sim 10^{-44}$ GeV, we obtain an upper bound on $M$:
\begin{equation}
M \lsim 10^{10} (\alpha\kappa)^{-1/2}\; {\rm GeV}.
\end{equation}

Another potential constraint outside the solar system comes from the period change of binary pulsar systems. Note that even if the velocities of the pulsars are smaller than $c_s$, the Goldstone boson can still be emitted through the quadrupole radiation.  To simplify the problem, we assume that two pulsars of equal mass $M_0$ move in a circular orbit separate by a distance $2r_0$ with an angular frequency $\omega_0$. The angular frequency and the orbital velocity $v_0$ are related by $v_0= \omega_0 r_0$. We also ignore the velocity of the center of the system.

The energy loss due to the multipole radiation in the case where $v_0<c_s$ or
\v{C}erenkov radiation in the case where $v_0> c_s$ can be calculated in the same way
described in App.~\ref{sec:simpleway}.
Analytic formulae can be obtained by taking different limits of the Bessel functions for
$v_0 \ll c_s$ or $v_0 \gg c_s$.

For $v_0 = \omega_0 r_0 \ll c_s$, the energy loss is dominated by the quadrupole radiation.
As derived in App.~\ref{sec:simpleway}, it is given by Eq.~(\ref{eq:case1}).
\begin{equation}
\dot{E}_{\rm src} = - \frac{4\alpha M_0^2 M^2 v_0^6}{15\pi M_{\rm Pl}^4 r_0^2 c_s^7} = -\frac{2^{12}\pi \alpha M^2 v_0^{10}}{15\, c_s^7}.
\end{equation}
On the other hand, the energy loss due to usual gravitational waves is given
by~\cite{Will:1981cz}
\begin{equation}
\dot{E}_{\rm tensor}= -\frac{1}{8\pi M_{\rm Pl}^2} \frac{1}{5}
\langle \dddot{Q}_{ij} \dddot{Q}_{ij} \rangle = -\frac{2}{5}\,
\frac{G_N^4 M_0^5}{r_0^5} = -\frac{2^{14}\pi M_{\rm Pl}^2 v_0^{10}}{5},
\end{equation}
where $Q_{ij}$'s are the quadrupole moments of the gravitational source.
Taking the ratio, we have
\begin{equation}
\frac{\dot{E}_{\rm src}}{\dot{E}_{\rm tensor}}
= \frac{\alpha M^2}{12 M_{\rm Pl}^2 c_s^7}
= \frac{\alpha_2^{\rm PPN}}{6\,c_s^3} \simeq 10^2
\left(\frac{\alpha_2^{\rm PPN}}{4\times 10^{-7}}\right)
\left(\frac{10^{-3}}{v_0}\right)^3 \left(\frac{v_0}{c_s}\right)^3.
\end{equation}
The observed period change of the binary pulsars PSR 1913+16 is consistent with the energy
loss entirely due to the gravitational wave radiation~\cite{Will:1981cz}, so the above ratio
must be smaller than 1.
Given their orbital velocity $v_0 \sim 10^{-3}$, we can see that for $v_0 \ll c_s$
(say, $c_s > 10\, v_0$), this does not give a stronger bound than the
$\alpha_2^{\rm PPN}$ bound from the solar system.
For $v_0$ getting close to $c_s$, this can compete with the solar system bound,
but our approximation starts to break down.

Next, let us consider the opposite limit, $c_s \ll v_0$. The energy loss is given approximately by  (see  Eq.\ (\ref{eq:case2}) of App.~\ref{sec:simpleway})
\begin{equation}
\label{eq:large_v0}
\dot{E}_{\rm src} \simeq -\frac{\alpha M^2 M_0^2}{4\pi M_{\rm Pl}^4 c_s r_0^2} \sum_{n=1}^{\infty} \frac{2n v_0}{c_s} \left| f\left(\frac{2n v_0}{c_s r_0}\right)^2 \right|.
\end{equation}
The problem is that the arguments in the form factor take values much bigger than $1/r_0$ in this case. As we discussed before, nonlinear effects are already important at that length scale and the region of the ether being dragged by the pulsars is approximately of size $r_0$. The effective form factor must be very suppressed and is difficult to calculate from first principles. Physically, the binary system emits ether waves with angular frequencies that are integer multiples of $2\omega_0$. Because the sound velocity of the ether is much smaller than the orbital velocity, the wavelengths of the ether waves are much shorter than the size of the system. Therefore, the amplitudes of the ether waves must be highly suppressed. If we assume a power law suppression that $f(2nv_0/c_s r_0) \sim (2nv_0/c_s)^{-p}$ with $p>1$, then the sum in \Eq{large_v0} is $\sim (c_s/v_0)^{2p-1}$, and the energy loss would be
\begin{equation}
\dot{E}_{\rm src} \sim -\frac{\alpha M^2 M_0^2}{4\pi M_{\rm Pl}^4 r_0^2 c_s} \left(\frac{c_s}{v_0}\right)^{2p-1} \sim 10^3 \alpha \left(\frac{M}{10^{10}\; {\rm GeV}}\right) \left(\frac{10^{-3}}{v_0}\right)^7 \left(\frac{c_s}{v_0}\right)^{2p-2} \frac{dE_{\rm tensor}}{dt}.
\end{equation}
Obviously for $c_s \ll v_0$ this does not give a significant constraint. For $c_s$ getting close to $v_0$ it may become a competitive constraint depending on the power of the suppression factor, but again our approximation starts break down.

\section{Couplings to the Standard Model}
\label{sec:sminteractions}

Given that gauged ghost condensation yields such a modest modification of gravity, the strongest bounds on this model of Lorentz breaking could come from direct couplings between $\scr{A}_\mu$ and the standard model.  As we will see, most couplings could be forbidden by discrete symmetries, 
but we will consider what would happen if the gauged ghost condensate were not confined to a hidden sector.

Even in the absence of gravitational couplings, the Lagrangian in \Eq{Leff} is fascinating because it is the effective field theory description of an Abelian gauge theory in a Lorentz-violating Higgs phase. To our knowledge, the dynamics of this phase has not been studied in the literature. Like the Lorentz-invariant Higgs phase, the gauge boson eats a Goldstone and therefore has three polarizations, but what is especially bizarre about this system is that all three polarizations are massless.   There are long-range ``magnetic'' interactions in this theory but ``electric'' interactions are exponentially suppressed.   In other words, in order to produce a healthy Newton's law for gravity, we had to sacrifice Coulomb's law for this ghost-electromagnetic $U(1)$.

Note that this $U(1)_{\rm ghost}$ theory is different from other Lorentz-violating $U(1)$ gauge theories considered in the literature.  In theories based on \Ref{Colladay:1998fq}, the Lagrangian contains gauge invariant but Lorentz-violating terms like $k_{\alpha \beta \gamma \delta} F^{\alpha \beta} F^{\gamma \delta}$, where $k_{\alpha \beta \gamma \delta}$ is some fixed Lorentz-violating tensor.  Because gauge symmetry is maintained, the equations of motion for electric and magnetic fields are modified without introducing new propagating degrees of freedom.  A model that does violate gauge symmetry is given in \Ref{Dvali:2005nt} which considers a ``magnetic'' Higgs phase where Coulomb's law is virtually unchanged.  Starting with the proposal of \Ref{Bjorken:1963vg}, there has also been speculation that the electromagnetic field could arise as Goldstone bosons from spontaneous Lorentz breaking.  In contrast to these other theories, gauged ghost condensation is a model where 
 a $U(1)$ gauge field enters an ``electric'' Higgs phase triggered by a Lorentz-violating vev for a charged scalar field.

\subsection{Catalog of Allowed Couplings}

At minimum, even if we do not include direct interactions between standard model fields and $\scr{A}_\mu$, we expect graviton loops to generate interactions of the form
\beq[nowaytostopcoupling]
\Delta\scr{L} \sim c \frac{M^2}{\MP^4}T^{\mu\nu}\scr{A}_\mu \scr{A}_\nu,
\eeq
where $T_{\mu\nu}$ is a symmetric dimension four standard model operator, and $c$ is expected to be $\scr{O}(1)$.  Note that there is no constraint from gauge invariant on this coupling because $\scr{A_\mu} =A_\mu +M\d_\mu \phi$ is a $U(1)_{\rm ghost}$ gauge invariant combination.
The $1/\MP^4$ suppression arises because $\scr{A}_\mu$ contains $\partial_\mu \phi$ so this interaction is actually dimension eight, and the factors of $M$ are inserted because $\scr{A}_\mu$ is defined with mass dimension $+1$. Setting $\scr{A}_\mu$ to its vacuum expectation value $\avg{\scr{A}_0} = M$:
\beq
\Delta\scr{L} \sim c \frac{M^4}{\MP^4}T^{00}.
\eeq
If we take $T^{\mu\nu}$ to be the stress-energy tensor for a standard model fermion, the inclusion of $T^{00}$ allows the maximal attainable velocity of the fermion to differ from the speed of light:
\beq
\delta v_\psi \sim \frac{M^4}{\MP^4}.
\eeq
Constraints from high precision spectroscopy and the absence of vacuum \v{C}erenkov radiation bound $\delta v_\psi \lsim 10^{-21} - 10^{-23}$ \cite{Coleman:1998ti}.   This limits the scale of spontaneous Lorentz violation to be $M \lsim 10^{13} \GeV$.  Remarkably, for $g \sim 1/10$ this bound on $M$ is comparable to the bound from the gravitational sector in Sec.~\ref{sec:togravity}.

If there is a $\scr{A}_\mu \rightarrow - \scr{A}_\mu$ symmetry, then \Eq{nowaytostopcoupling} is the leading coupling between the gauged ghost condensate and the standard model.  Relaxing this symmetry allows couplings of the form
\beq
\Delta \scr{L} = \tilde{q} J^\mu \scr{A}_\mu,
\eeq
where $J^\mu$ is a standard model current and $\tilde{q}$ is the coupling constant. In fact, this is
just equivalent to charging standard model fields under
the $U(1)_{\rm ghost}$ gauge symmetry. To see this, consider the interaction
\beq[vectorcoupling]
\tilde{q}_{\psi}\,\scr{A}_\mu \bar{\psi}\gamma^\mu \psi 
=\tilde{q}_{\psi}\,({A}_\mu+M\d_\mu \phi) \bar{\psi}\gamma^\mu \psi,
\eeq
where $\psi$ is a standard model fermion. If the current $\bar{\psi}\gamma^\mu \psi$ is conserved, its interaction with
$\d_\mu \phi$ can be removed by a field redefinition
\beq
\psi \to \psi' = e^{-i \tilde{q}_\psi M\phi } \psi
\eeq
into the kinetic term. In terms of the redefined field $\psi'$,
the interaction \Eq{vectorcoupling} becomes
\beq
\tilde{q}_{\psi} A_\mu \bar{\psi}' \gamma^\mu \psi' \,
\eeq
which is just a $U(1)_{\rm ghost}$ gauge interaction of a fermion with 
charge $\tilde{q}_{\psi}$. The point is that $\exp(-iM\phi)$ carries a unit charge 
of the $U(1)_{\rm ghost}$ gauge symmetry, so the charge of a field under 
$U(1)_{\rm ghost}$ can be altered
by multiplying it by powers of $\exp(-iM\phi)$. It is most convenient,
however, to work in the $\psi'$ basis where the $\d_\mu \phi$ coupling is
removed.

New long-range forces other than electromagnetism and gravity are
strongly constrained. In gauged ghost condensation the $\scr{A}_0$ field
is massive, so there is no $1/r$ potential between $U(1)_{\rm ghost}$
charges in the preferred rest frame. Nonetheless, $\scr{A}_i$ is still 
massless, giving rise to additional magnetic forces. Effectively, this
gives a tree-level contribution to the magnetic moment of a particle
charged under the $U(1)_{\rm ghost}$. If we assume that the $U(1)_{\rm ghost}$
charges of the standard model fields are proportional to their $U(1)_{\rm EM}$
charges, the correction to the Land\'{e} $g$ factor is
\beq[gfactorchange]
\delta \left(\frac{g-2}{2} \right) \sim 
\frac{\tilde{q}_{e}^2\alpha_{\rm ghost}}{\alpha_{\rm EM}},
\eeq
where $\alpha_{\rm ghost}=g^2/(4\pi)$. 
Note that in theories with a 
normal para-$U(1)$ gauge group that ``regauges'' 
electromagnetism~\cite{Holdom:1985ag}, 
this correction is absent, because the para-$U(1)$ also shifts the 
electrostatic charges of the fermions. The most accurately measured $g-2$
is for the electron with an uncertainty at the $10^{-9}$ level \cite{VanDyck:1987ay, CODATA}.  However,
since it serves as a definition of $\alpha_{\rm EM}$, to constrain
$\tilde{q}_e g$ we need the next most precise determination 
of $\alpha_{\rm EM}$.
A measurement of $\alpha_{\rm EM}$ using the atom interferometry with
the laser-cooled cesium atoms has reached the $10^{-8}$ uncertainty level
and the result is in agreement with the $g-2$ measurement~\cite{Chu}.
This implies a constraint
\beq[gminustwoconstraint]
 \frac{\tilde{q}_e^2\alpha_{\rm ghost}}{\alpha_{\rm EM}} \lsim 10^{-8}  
\qquad \Longrightarrow \qquad \tilde{q}_e g \lsim 10^{-4}.
\eeq
As we will see, this bound is much weaker than the bound from the long-range force
derived in the next subsection: for sources moving with respect to
the preferred frame, there is a direction-dependent pseudo-Coulomb potential.
If the $U(1)_{\rm ghost}$ charges are not proportional to the electromagnetic 
charges, then the measurement of $(g-2)$ would depend on which species of 
fermion was responsible for setting up the background magnetic field, 
though in practice, almost all laboratory magnetic fields are in some way 
established by electrons.

One can also consider the couplings of standard model antisymmetric 
tensors to the field tensor of the $U(1)_{\rm ghost}$. At the lowest order,
we can write down a kinetic mixing between the $U(1)_{\rm ghost}$ and
$U(1)_{\rm EM}$,
\beq
\Delta \scr{L} = c_o F_{\mu\nu}  F_{\rm EM}^{\mu\nu}.
\eeq
Such a term can be removed by a redefinition of the $U(1)$ gauge fields,
resulting in a shift of the $U(1)_{\rm ghost}$ charges of the standard
model fields, so the effect is the same as discussed in the previous
paragraph. Alternatively, we could imagine dipole interactions
\beq[bogdancoupling]
\Delta \scr{L} = \frac{1}{\Lambda} \sum_\Psi c_\Psi \bar{\Psi} \sigma^{\mu\nu} \Psi F_{\mu\nu}.
\eeq
These couplings were studied in a Lorentz invariant setting in \Ref{Dobrescu:2004wz}, where the bound on $\Lambda$ for the electron was expressed in terms of the Higgs vev $\avg{h} = 174 \GeV$:
\beq
\frac{\Lambda}{c_e} > \frac{(3 - 60 \TeV)^2}{\avg{h}} \sim 10^5 - 10^7 \GeV,
\eeq
where the more stringent bound comes if \Eq{bogdancoupling} introduces flavor mixing between electrons and muons.  It would be interesting to study these couplings more carefully in the Lorentz-violating Higgs phase, but we do not expect significant departures from these bounds.

Because we will find strong bounds on $\tilde{q}_\psi$ from direction dependent pseudo-Coulomb potentials, we want to forbid the coupling in \Eq{vectorcoupling} while still allowing new couplings other than just  \Eq{nowaytostopcoupling}.  One way to forbid a coupling to the electromagnetic current is to choose $\phi$ to be parity-odd, or equivalently
\beq[paritytransform]
P: \qquad \scr{A}_0 \rightarrow -\scr{A}_0, \qquad \scr{A}_i \rightarrow \scr{A}_i.
\eeq  
In the standard model, electroweak interactions violate parity, and because we expect the scale $M$ to be much higher than the electroweak scale, at best we can say that the coupling of $\scr{A}_\mu$ to parity-even currents should be suppressed relative to parity-odd currents.   For a fermion $\Psi$ with Dirac mass $m_D$, we can construct the vector and axial currents:
\beq
J_\mu = \bar{\Psi} \gamma_\mu \Psi, \qquad J^5_\mu = \bar{\Psi} \gamma_\mu \gamma^5 \Psi.
\eeq
By looking at the relevant triangle diagram, we estimate that the coupling $\scr{A}^\mu J_\mu$ should be suppressed relative to the coupling $\scr{A}^\mu J^5_\mu$ by $m^2_D G_F/16\pi^2$, where $G_F$ is Fermi's constant.  For an electron, this suppression is $10^{-14}$, so to a good approximation we can ignore couplings to parity-even currents.

With the above caveats, the leading coupling of $\scr{A}_\mu$ to any light Dirac fermion, in particular electrons and nucleons, is
\beq[spincoupling]
\scr{L}_{\rm int} = \frac{M}{F} \scr{A}^\mu J^5_\mu,
\eeq
where we parametrize the coefficient as a ratio of two scales $M$ and $F$
to compare with the result of ungauged ghost condensation.
Setting $\scr{A}_\mu$ to its vev:
\beq[cptoddpiece]
\scr{L}_{\rm int} = \mu \bar{\Psi} \gamma^0 \gamma^5 \Psi, \qquad \mu = \frac{M^2}{F}.
\eeq
This Lorentz- and CPT-violating term is the same as in ghost condensation \cite{Arkani-Hamed:2003uy,Arkani-Hamed:2004ar}, and gives rise to different dispersion relations for the left- and right-handed fermions \cite{Kostelecky:1999zh,Andrianov:2001zj}.  In a frame boosted with respect to the preferred frame, \Eq{cptoddpiece} contains the interaction
\beq
\mu \vec{s} \cdot \vec{v}_{\rm earth},
\eeq
where we have identified $\bar{\Psi} \vec{\gamma} \gamma^5 \Psi$ with non-relativistic spin density $\vec{s}$.  Assuming the preferred rest frame is aligned with the CMBR, we take $|\vec{v}_{\rm earth}| \sim 10^{-3}$, and experimental bounds on velocity-dependent spin couplings to electrons is $\mu \sim 10^{-25} \GeV$ \cite{Heckel:1999sy} and to nucleons $\mu \sim 10^{-24} \GeV$ \cite{Phillips:2000dr,Cane:2003wp}.

After integrating out $\scr{A}_0$, we are left with the interaction
\beq[axialcoupling]
\scr{L}_{\rm int} = \frac{M}{F} \vec{s} \cdot \vec{\scr{A}}.
\eeq
The $g \rightarrow 0$ limit of this coupling was discussed in \Ref{Arkani-Hamed:2004ar} and it led to two Lorentz-violating dynamical effects:  ``ether'' Cerenkov radiation and a long-range spin-dependent potential.
These effects in this model will be discussed in the next subsection.

\subsection{Dynamics of the Lorentz-Violating Higgs Phase}

The Lagrangian in \Eq{Lgaugedghostlin} is the most general Lagrangian with an $\scr{A}_\mu \rightarrow -\scr{A}_\mu$ symmetry that is spontaneously broken by the vacuum $\avg{\scr{A}_0} = M$.   If we only enforce \Eq{paritytransform}, then we can add additional couplings to the gauged ghost Lagrangian such as
\beq
\Delta \scr{L} \sim M (\scr{A}_0 - M) \partial_i \scr{A}^i.
\eeq
When we integrate out $\scr{A}_0$, however, this term does not change the basic form of \Eq{Leff}, and the effective Lagrangian for $\scr{A}_i$ at distances large compared to $1/M$ is
\beq
\scr{L}_{\rm eff} = \frac{1}{2c_v^2g^2} (\d_t \vec{\scr{A}})^2
- \frac{1}{4 g^2} F_{ij}^2
- \sfrac 12 \al (\grad \cdot \vec{\scr{A}})^2,
\eeq
where we have absorbed the effect of $\beta_1$ and $\alpha_2$ into the parameters $g$ and $c_v$, and $c_v$ has the interpretation of the velocity of the transverse vector modes.

Going to $\scr{A}_i \equiv A_i$ gauge, it is straightforward to calculate the $A_i A_j$ propagator,
\beq[aaprop]
\langle A_i A_j \rangle = \frac{1}{(\omega / c_v)^2 - k^2}\left(g^2 \delta_{ij} +   (g^2 - 1/\alpha)\frac{k_i k_j}{(\omega/c_s)^2 - k^2} \right),
\eeq
where $c_s = c_v g \sqrt{\alpha}$ is the velocity of the longitudinal mode of $A_i$.  We will always assume that $c_s \ll c_v$.  The first thing to check is what kind of pseudo-Coulomb potential is generated between two particles carrying $U(1)_{\rm ghost}$ charges in a frame moving with respect to the preferred frame.  Consider a source of charge $\tilde{Q}$ under $U(1)_{\rm ghost}$ moving with velocity $v < c_s$:
\beq
\bal
J^0 &= \tilde{Q} \delta^{(3)}(\vec{x} - \vec{v}t) \\
\vec{J} & = \tilde{Q} \vec{v} \delta^{(3)}(\vec{x} - \vec{v}t)
\eal
\qquad
\bal
\tilde{J}^0 &= 2 \pi \tilde{Q} \delta(\omega - \vec{k}\cdot\vec{v}) \\
\vec{\tilde{J}} & = 2\pi \tilde{Q} \vec{v} \delta(\omega - \vec{k}\cdot\vec{v}) 
\eal
\eeq
where there are corrections at $\scr{O}(v^2)$.  To this order, we can use the $\omega \rightarrow 0$ limit of \Eq{aaprop}.   Taking the Fourier transform of the propagator, the potential between $\tilde{Q}$ and a test charge $\tilde{q}$ a comoving distance $r$ away is
\beq[angularcharge]
V(r,\theta) = -\frac{{g}^2 v^2 \tilde{Q} \tilde{q}}{8\pi r}\left(1+ \cos^2\theta + \frac{1}{g^2 \alpha} \sin^2 \theta \right),
\eeq
where $\cos \theta= \hat{r}\cdot \hat{v}$.  In the limit $c_s \ll c_v$, the third term dominates, yielding an effective anomalous gauge coupling
\beq
g_{\rm anom} \sim \frac{ v }{\sqrt{\alpha}},
\eeq
and angular dependence with respect to the velocity of the preferred frame.  
Bounds on new gauge interactions are usually derived by noting the absence of any long-range forces other than electromagnetism and gravity.  In particular, precision tests of gravity using materials with different baryon-to-lepton ratios place limits on a possible $U(1)_{B-L}$ gauge coupling of $g_{B-L} < 10^{-23}$ \cite{Adelberger:2003zx}.
If searches for $U(1)_{B-L}$ gauge couplings are sensitive to such velocity effects, then, $ \tilde{q} g_{\rm anom} \lsim 10^{-23}$. 
Assuming the earth is moving with velocity $|\vec{v}| \sim 10^{-3}$ with respect to the preferred rest frame, then
\beq[tildegbound]
{\tilde{q}_\psi} \lsim 10^{-20} \sqrt{\alpha},
\eeq
where $\psi$ represents an ordinary particle $p$, $n$, or $e$.   For reasonable values of $\alpha$ and $g$, this bound is stronger than \Eq{gminustwoconstraint}, indicating that couplings between $\scr{A}_\mu$ and vector currents must be strongly suppressed.  Still, it would be interesting to search for the angular dependent Coulomb potential in \Eq{angularcharge} to see whether the bound on ${\tilde{q}_\psi}$ could be improved.

Even if we forbid the coupling to vector currents using \Eq{paritytransform}, axial couplings are still allowed.  The coupling in \Eq{spincoupling} between $\vec{A}$ and spin is familiar from ghost condensation and gives rise to a spin-dependent inverse-square law force \cite{Arkani-Hamed:2004ar}.  Consider spins $\vec{S}_1$ and $\vec{S}_2$ separated by a distance $r$ that are at rest with respect to the preferred frame.   Taking the Fourier transform of \Eq{aaprop} with $\omega \rightarrow 0$, the $1/r$ potential between them takes the form
\beq[novelocitypotential]
V(r) = -\frac{M^2}{F^2}\frac{1}{8\pi r}\left((g^2 + 1/\alpha)\vec{S}_1 \cdot \vec{S}_2 + (g^2-1/\alpha)(\vec{S}_1 \cdot \hat{r})(\vec{S}_2 \cdot \hat{r})   \right).
\eeq
This reduces to the pure Goldstone result in the $g \rightarrow 0$ limit, modulo a redefinition $M \rightarrow \sqrt{\alpha}M$.  This potential has an interesting feature which distinguishes it from both electromagnetism and ghost condensation.  Consider a toroidal solenoid filled with a ferromagnetic material.\footnote{Thanks to Blayne Heckel for suggesting this geometry.}   When current runs through the solenoid, the ferromagnetic spins will align in the azimuthal direction, but as we will show in App.~\ref{ringofspin}, there will be no net magnetic field nor net longitudinal $\vec{A}$ field outside of the solenoid.  However, there will be transverse $\vec{A}$ fields which can interact with a test spin, and because there is no magnetic field leakage from this geometry, this allows for a null test for the inverse-square law spin-dependent force.

In order for the potential in \Eq{novelocitypotential} to be experimentally testable, it has to be at least roughly of gravitational strength, and if we assume spin sources with one aligned spin per nucleon, $M/F$ has to be of order $m_N/\MP \sim 10^{-19}$.  As shown in \Ref{Arkani-Hamed:2004ar}, $M/F$ cannot be much larger than $10^{-19}$ without the theory exiting its range of validity.  When $M/F$ is so small, it is easy to satisfy the condition in \Eq{tildegbound}.  In the previous section, we argued that the ratio of the vector coupling to the axial coupling can be made as small as $m_e^2 \MP/16\pi^2 \sim 10^{-14}$ without fine tuning if we assume the quasi-parity invariance in \Eq{paritytransform}.  This tells us that
\beq
\frac{\tilde{q}_\psi}{M/F} < 10^{-14} \qquad \Longrightarrow \qquad \tilde{q}_\psi < 10^{-35} 
\eeq
in agreement with \Eq{tildegbound}.

This spin-spin potential is even more interesting at finite velocity.  For sources moving at velocity $\vec{v}$ will respect to the preferred frame, the potential at fixed comoving distance is given by the Fourier transform of \Eq{aaprop} with the replacement $\omega \rightarrow \vec{k} \cdot \vec{v}$.
In the case $v < c_s$, we can calculate the spin-spin potential as a power series in $v/c_s$.  For simplicity, we define
\beq
f(\vec{A}, \vec{B}) =\vec{A}\cdot\vec{B} - (\vec{A}\cdot\hat{r})(\vec{B} \cdot\hat{r}),
\eeq
and the potential takes the form
\beq
\bal
\frac{F^2}{M^2}V(r) & =- \frac{g^2}{4\pi r}\vec{S}_1 \cdot \vec{S}_2 - \frac{1/\alpha - g^2}{8\pi r}f(\vec{S}_1, \vec{S}_2) \\
&\quad -  \frac{1/\alpha - g^2}{32\pi r}\frac{v^2}{c^2_s}\left(f(\vec{S}_1, \vec{S}_2)f(\hat{v}, \hat{v}) + 2f(\vec{S}_1, \hat{v})f(\vec{S}_2, \hat{v})  \right) + \scr{O}(v^4/c^4_s).
\eal
\eeq
The unique angular dependence of this spin-spin potential is a smoking gun for gauged ghost condensation.  As $v$ gets larger than $c_s$, the potential should map onto the shadow/shockwave form of \Ref{Arkani-Hamed:2004ar}, though to see this behavior, one would need to keep track of $k^4$ terms in the dispersion relation for the $\scr{A}_i$ longitudinal mode.

Next, we consider energy loss due to the emission of $\scr{A}_i$ particles.
In \Ref{Arkani-Hamed:2004ar} we did not consider bounds on new light particles from astrophysics \cite{Raffelt}, because the scale of spontaneous Lorentz violation $M$ was constrained to be much lower than typical astrophysical energies.  Now that we can raise the scale of Lorentz violation well above even the electroweak scale, these bounds are relevant, but its easy to see that once we impose the experimental bound on $\mu$, the astrophysical bound is almost automatically satisfied.  For a pseudo-scalar $\varphi$ coupling to the axial current, stellar energy loss arguments bound the coefficient of the operator
\beq[axionbound]
\frac{1}{F}J_\mu^5 \partial^\mu \varphi
\eeq
to be $F > 10^9 \GeV$ \cite{Raffelt:1999tx}.  In the non-relativistic limit, \Eq{axionbound} is identical to the canonically normalized coupling of the longitudinal mode of $\vec{\scr{A}}$ to spin-density, so as long as $M \gsim 10 \eV$, the bound on $\mu$ is stricter than the bound on $F$ directly.  For the transverse modes of $\vec{\scr{A}}$, we can estimate the bound on $M/F$ by going to canonical normalization and replacing the derivative $\partial^\mu$ with a typical astrophysical energy $E_{\rm astro} \sim 100 \keV$:
\beq
\frac{1}{F}J_\mu^5 \partial^\mu \varphi \sim \frac{E_{\rm astro}}{F} \vec{J}^{\,5} \cdot \vec{A}^c \qquad  \Longrightarrow \qquad \frac{M}{F}g < 10^{-13}.
\eeq
As long as $M \gsim10^{-3} \eV / g$, the bound on $\mu$ enforces the astrophysical bound. This bound also applies to the vector coupling, requiring
\beq
\tilde{q}_\psi g < 10^{-13}.
\eeq

Finally, for sources moving with respect to the preferred frame, there is both the possibility of \v{C}erenkov radiation from gravitational couplings (see Sec.~\ref{sec:Cerenkov}) and from the coupling in \Eq{cptoddpiece}.%
\footnote{Note that there is no \v{C}erenkov radiation from the coupling in \Eq{vectorcoupling}, because it only involves couplings to the transverse modes of $A_i$, which travel at $c_v$, \emph{i.e.}\ nearly the speed of light.}
  Because the scalar mode in gauged ghost condensation has a finite velocity, this \v{C}erenkov effect is turned off when the velocity of the source $v$ is smaller than $c_s$.  
When $v/c_s \gg 1$, we expect the discussion of \Ref{Arkani-Hamed:2004ar} to go through virtually unchanged, where we found that within the range of validity of the effective theory, there were no observable effect from ether \v{C}erenkov radiation for electrons.   \Ref{Grossman:2005ej} studied the effect of \v{C}erenkov radiation on neutrinos from SN1987A, where the bound was expressed as:
\beq
\left(\frac{M}{F} \right)_\nu < \frac{10^{-16}}{g^{3/2}}.
\eeq
Assuming that neutrino couplings are comparable to electron couplings, this does not place any additional bounds on couplings between $\scr{A}_\mu$ and axial currents.

\section{Conclusions}
\label{sec:conclusions}
We have analyzed spontaneous breaking of Lorentz symmetry by a vector condensate,
a theory which gives rise to 3 extra modes in the gravitational sector.
We showed that this theory can be viewed as a gauging of ghost condensation,
the minimal theory of spontaneous breaking of Lorentz symmetry.
The relation to ghost condensation allows us to understand the power counting
of operators in the effective theory, and also to understand the limit where
the gauge coupling $g$ is small, where we recover ghost condensation.
We also analyzed various constraints on this model, taking into account
important nonlinear effects that are closely analogous to those found
in ghost condensation \cite{Arkani-Hamed:2004ar}.

We find that the limits on the scale $M$ where Lorentz symmetry is
spontaneously broken allow values much higher than the weak scale,
suggesting that Lorentz violation may be more closely connected with
fundamental particle physics.
From purely gravitational constraints, we find that
$M \lsim \mathrm{Min} (10^{12} \GeV,\, g^2 \, 10^{15} \GeV)$ 
from observation of a steller-mass black hole and constraints on the PPN
parameters. 
If we allow all possible couplings to the standard model and
cut off gravity loop effects at the scale $M$, we find that we need
$M \lsim 10^{13} \GeV$ from the absence of observed Lorentz violation.

It is our hope that this work will bring Lorentz violation closer
to the mainstream of particle physics, and that it will stimulate
further investigations on the possible relation to the puzzles of
gravity and cosmology.

\setcounter{section}{0}
\renewcommand{\thesection}{\Alph{section}}

\section{Decoupling Limit and Comparison with the Literature}
\label{app:decoupling}

Alternative theories of gravity with additional vector fields pointing
in some preferred direction have been considered in the 
literature~\cite{Will:1972,Hellings:1973,Jacobson:2001yj,Eling:2003rd,Jacobson:2004ts,Kostelecky:1989jw,Kostelecky:2003fs,Clayton:1998hv,Clayton:2001vy,Moffat:2002nm,Gripaios:2004ms}.
The value of the vector field is often imposed as a constraint through
a Lagrange multiplier
\beq[langrangeconst]
\scr{L} \supset -\kappa (u^\mu u_\mu -1).
\eeq
We will show that this corresponds to certain decoupling limit of our theory.
To see this, we can add a term $-2\kappa^2/M^4$ to the Lagrange
multiplier. Integrating out $\kappa$ we obtain
\beq
\mathcal{L} \supset \frac{M^4}{8}(u^2-1)^2
\eeq
which is exactly the leading term of our gauged ghost condensate Lagrangian:
\beq
\mathcal{L} \subset \frac{M^4}{8}\left(\frac{\scr{A}^\mu \scr{A}_\mu}{M^2} - 1\right)^2,
\eeq
making the appropriate mapping between $u^\mu$ and $\scr{A}_\mu$.  The constraint of the Lagrange multiplier is obtained by sending
$M \to \infty$. Therefore, we see that these theories correspond
to the decoupling limit $M \to \infty$ of gauged ghost condensation,
with appropriate rescaling of other parameters, $g^{-2}, \alpha,
\beta, \gamma,\ldots \to 0$ while keeping $g^{-2}M^2, \alpha M^2,
\beta M^2, \gamma M^2,\ldots$ finite.

In this appendix we will study this decoupling limit and see the extent to which they are healthy modifications of gravity.
We start with 
the ungauged ghost condensate with the lagrangain
\beq
\scr{L} = \scr{L}_{\rm EH}
+ \sfrac 18 M^4 (X - 1)^2 - \sfrac 12 \tilde{M}^2 (\grad^2 \pi)^2,
\eeq
where $\scr{L}_{\rm EH}$ is the Einstein-Hilbert Lagrangian.
For convenience we have defined $\tilde{M}^2 = \alpha M^2$.
We want to understand the physics of taking $M \to \infty$, with
$\tilde{M}$ and $\MP$ held fixed.
This is equivalent to imposing the constraint $X = 1$.

The linearized Lagrangian for the scalar sector including a matter source $\rho$ is
\beq
\scr{L} = - \MP^2 (\grad \Phi)^2 + \sfrac 12 M^4 (\Phi - \dot\pi)^2
- \sfrac 12 \tilde{M}^2 (\grad^2 \pi)^2 + \rho \Phi.
\eeq
For large $M$, we impose the constraint $\Phi = \dot\pi$, which
gives the effective Lagrangian
\beq
\scr{L}_{\rm eff} = -\MP^2 (\grad \dot\pi)^2 
- \sfrac 12 \tilde{M}^2 (\grad^2 \pi)^2 + \rho \dot \pi.
\eeq
From this we easily read off the dispersion relation for the
scalar mode:
\beq
\om^2 = - \frac{\tilde{M}^2}{2 \MP^2} \vec{k}^2.
\eeq
Note that this has the wrong sign for $\tilde{M}^2 > 0$,
which is just a manifestation of the Jeans instability in ghost condensation.
But for $\tilde{M}^2 < 0$, the scalar has a healthy dispersion
relation.

Of course, the same results can be obtained from the dispersion
relation for $\pi$ with finite $M$, 
Rewriting in terms of $\tilde{M}$:
\beq
\om^2 = - \frac{\tilde{M}^2}{2\MP^2} \vec{k}^2
+ \frac{\tilde{M}^2}{M^4} \vec{k}^4.
\eeq
In the limit $M \to \infty$ only the first term survives.
So the decoupling limit is equivalent to considering
\beq
|\vec{k}| \ll m = \frac{M^2}{\sqrt{2} \MP}.
\eeq
Therefore, we can define a healthy theory as long as
$m$ is large enough to be a UV scale and if we change the sign of
$\tilde{M}^2$ relative to the ghost condensate.  

What is the Newtonian potential in this limit? Define the decoupling limit sound speed as
\beq
v_0^2 = - \frac{\tilde{M}^2}{2\MP^2}.
\eeq
In the $|\vec{k}| \ll m$ limit, the $\Phi\Phi$ propagator is
\beq
\avg{\Phi \Phi} \to -\frac{1}{2 \MP^2} \frac{1}{\vec{k}^2} \left( \frac{\omega^2}{\omega^2 - \vec{k}^2 v_0^2}\right).
\eeq
At very late times ($\omega \rightarrow 0$), the theory does not have a Newtonian potential, so the decoupling limit of ghost condensation is not a viable modification of gravity at very late times.  However, for times $t$ short compared to $r / v_0$, $|\vec{k}| \gg \omega \gg v_0 |\vec{k}|$
\beq
\avg{\Phi \Phi} \to -\frac{1}{2 \MP^2} \frac{1}{\vec{k}^2} + \mathcal{O} \left( \frac{\vec{k}^2 v_0^2}{\omega^2}   \right),
\eeq
and we recover an ordinary Newtonian potential plus small corrections.  If $v_0$ is sufficiently small, then the decoupling limit could be a viable modification of gravity with interesting late time behavior, and we will study this limit in future work.

Now consider the gauged ghost condensate.
To make the limit clearer, we define $\pi$ to have mass dimension
$-1$, $A_\mu$ to have mass dimension $+1$, and rescale the fields and
couplings relative to \Eq{Lgaugedgravghostlin} by
\beq
A_\mu \to \frac{M}{\mu} A_\mu, \quad \frac{1}{g^2} \to 
\left(\frac{\mu}{M}\right)^2 \frac{1}{g^2},
\eeq
where $\mu$ is a mass scale held fixed when we take the limit $M \to \infty$.
The gauge transformations can now be written as
\beq
\de A_\mu = \d_\mu \chi,
\qquad
\de \pi = -\frac{\chi}{\mu},
\eeq
and the scalar sector Lagrangian is
\beq
\scr{L} = -\MP^2 (\grad\Phi)^2 + \frac{1}{2 g^2} (\grad A_0)^2
+ \sfrac 12 M^4 (\Phi - \dot\pi + A_0 / \mu)^2
- \sfrac 12 \tilde{M}^2 (\grad^2 \pi)^2.
\eeq
We are
interested in the limit $M \to \infty$ with the other couplings
(including $\mu$) held fixed.
The dispersion relation for the scalar mode is
\beq
\om^2 = -\frac{\tilde{M}^2}{2 \MP^2} \left( 1 - \frac{2 g^2 \MP^2}{\mu^2} \right)
\vec{k}^2 + \frac{\tilde{M}^2}{M^4} \vec{k}^4.
\eeq
As before, the second term is absent in the limit $M \to \infty$, and
is negligible for $|\vec{k}| \ll m$.
The $\vec{k}^2$ term has the right sign for
\beq
\tilde{M}^2 > 0,
\qquad
g > g_{\rm c} = \frac{\mu}{\sqrt{2} \MP},
\eeq
or for $\tilde{M}^2 < 0$, $g < g_{\rm c}$.
It is easy to see that in order to get a viable Newtonian potential at very late times ($\omega \rightarrow 0$), we must choose the first option.
The static limit of the $\Phi\Phi$ propagator is
\beq
\avg{\Phi\Phi} = -\frac{1}{2\MP^2} \, \frac{1}{\vec{k}^2}
\, \frac{1}{1 - g_{\rm c}^2 / g^2},
\eeq
so if $g < g_{\rm c}$ this would have the wrong sign relative to the standard Newtonian potential.  So as long as $g > g_c$, the decoupling limit of gauged ghost condensation is a healthy modification of gravity at leading order.  Interestingly, if $\tilde{M}^2 < 0$ and $g < g_{\rm c}$, then there is once again an intermediate range of times for which there is an ordinary Newtonian potential. 

This analysis shows why, for example, \Ref{Graesser:2005bg} finds finite corrections to $\alpha_2^{\rm PPN}$ in a model with the constraint in \Eq{langrangeconst}.  In Sec.~\ref{sec:togravity}, we found 
\beq
\alpha_{2}^{\rm PPN} \sim \frac{M^2}{\alpha g^4 \MP^2},
\eeq
which would diverge as $M^2 \rightarrow \infty$.  However in the decoupling limit where $g^{-2} M^2$ and $\alpha M^2$ are held fixed, $\alpha_{2}^{\rm PPN}$ stays finite.  We emphasize that while the decoupling limit is healthy, it obscures the power counting of gauged ghost condensation, which gives a systematic way to understand the relevance of operators in a Lorentz-violating setting.

\section{Stress-Energy Tensor of the Gauged Ghost Condensate}
\label{app:Tmunu}

The goal of this appendix is to calculate the stress-energy tensor of
the gauged ghost condensate. For this purpose we need to consider
general variations of the metric since the stress-energy tensor is
defined as the functional derivative of an action with respect to the
metric components. On the other hand, in order to make the expression 
independent of the heavy modes we have to integrate out the perturbation
of $D_0\phi$ before taking the variation and, thus, we need to separate
the time coordinate and space coordinates. For these reasons, we shall
adopt the ADM decomposition of the metric:
%
\begin{equation}
 ds^2 = -N^2dt^2 + q_{ij}(dx^i+\beta^idt)(dx^j+\beta^jdt)
\end{equation}
and define $\scr{A}_{\mu}$ by
%
\begin{equation}
 D_0\phi = (1+\scr{A}_0)N, \quad D_i\phi = \scr{A}_i. 
\end{equation}
We shall integrate $\scr{A}_0$ out.

Expanding $X-1$ as
%
\begin{equation}
 \frac{1}{2}(X-1) = \scr{A}_0 
  - \frac{1}{2}
  \left( q^{ij}-\frac{\beta^i\beta^j}{N^2} \right) \scr{A}_i\scr{A}_j 
  + O(\scr{A}_0^2),
\end{equation}
the leading Lagrangian is
%
\begin{equation}
 \scr{L} = \frac{M^4}{2}
  \left[
   \scr{A}_0 - \frac{1}{2}
   \left( q^{ij}-\frac{\beta^i\beta^j}{N^2} \right) \scr{A}_i\scr{A}_j 
       \right]^2 - \frac{M^2}{4g^2}F_{\mu\nu}F^{\mu\nu}
  -\frac{\alpha M^2}{2}(\nabla^{\mu}D_{\mu}\phi)^2,
\end{equation}
where $q^{ij}=(q^{-1})^{ij}$ and $q=\det q$. Integrating out the massive
mode $\scr{A}_0$,
%
\begin{equation}
 \scr{A}_0  = \frac{1}{2}
  \left( q^{ij}-\frac{\beta^i\beta^j}{N^2} \right) \scr{A}_i\scr{A}_j,
\end{equation}
we obtain
%
\begin{equation}
 F_{\mu\nu}F^{\mu\nu} = -2q^{ij}F_{\perp i}F_{\perp j}
  + q^{ik}q^{jl}F_{ij}F_{kl},
\end{equation}
and
%
\begin{equation}
 \nabla^{\mu}D_{\mu}\phi = -\frac{1}{\sqrt{q}}\partial_{\perp}
  \left\{\sqrt{q}
   \left[ 1-\frac{\beta^j\scr{A}_j}{N}
    +\frac{1}{2}\left(q^{kl}-\frac{\beta^k\beta^l}{N^2}\right)
    \scr{A}_k\scr{A}_l
       \right]\right\}
  + \frac{1}{N\sqrt{q}}\partial_i(N\sqrt{q}q^{ij}\scr{A}_j),
\end{equation}
where
%
\begin{equation}
 \partial_{\perp} \equiv \frac{1}{N}(\partial_t-\beta^i\partial_i),
\end{equation}
and
%
\begin{eqnarray}
 F_{\perp i} & \equiv & \frac{1}{N}(F_{ti}-\beta^kF_{ki}) \nonumber\\
  & = & \partial_{\perp}\scr{A}_i - \frac{1}{N}\partial_i
  \left\{ N
   \left[ 1+\frac{1}{2}\left(q^{kl}-\frac{\beta^k\beta^l}{N^2}\right) 
    \scr{A}_k\scr{A}_l\right]\right\}
  +\frac{\beta^k}{N}\partial_i\scr{A}_k,\\
  F_{ij} & = & \partial_i\scr{A}_j-\partial_j\scr{A}_i.
\end{eqnarray}

All components of the stress-energy tensor are given by taking the
variation of the effective action w.r.t. $N$, $\beta^i$ and $q_{ij}$. 
The result is
%
\begin{eqnarray}
 T_{\perp\perp} & \equiv &
  -\frac{1}{\sqrt{q}}\frac{\delta}{\delta N}
  \int dtd^3xN\sqrt{q}\scr{L}\nonumber\\
 & = & \frac{M^2}{g^2}
  \left\{\frac{1}{2}q^{ij}F_{\perp i}F_{\perp j}
   -\left[1+\frac{1}{2}\left(q^{kl}+\frac{\beta^k\beta^l}{N^2}\right)
     \scr{A}_k\scr{A}_l\right]
   \frac{1}{\sqrt{q}}\partial_i(\sqrt{q}q^{ij}F_{\perp j})
   + \frac{1}{4}q^{ik}q^{jl}F_{ij}F_{kl} \right\} \nonumber\\
 & & + \alpha M^2\left\{ -\frac{1}{2}(\nabla^{\mu}D_{\mu}\phi)^2
	      + \frac{\beta^i\scr{A}_i}{N}
	      \left(1+\frac{\beta^j\scr{A}_j}{N}\right)
	      \partial_{\perp}(\nabla^{\mu}D_{\mu}\phi)
	      -q^{ij}\scr{A}_i\partial_j(\nabla^{\mu}D_\mu\phi)\right\},
 \nonumber\\
 T_{\perp i} & \equiv & 
  -\frac{1}{\sqrt{q}}\frac{\delta}{\delta\beta^i}
  \int dtd^3xN\sqrt{q}\scr{L}
  \nonumber\\
 & = & \frac{M^2}{g^2}
  \left[ q^{jk}F_{ij}F_{\perp k} + \frac{\beta^j}{N}\scr{A}_i\scr{A}_j
   \frac{1}{\sqrt{q}}\partial_k(\sqrt{q}q^{kl}F_{\perp l})\right]
  \nonumber\\
 & & +\alpha M^2
  \left\{ -\left(1+\frac{\beta^j\scr{A}_j}{N}\right)\scr{A}_i
   \partial_{\perp}(\nabla^{\mu}D_{\mu}\phi) \right.\nonumber\\
 & & \left.
      -\left[1-\frac{\beta^j\scr{A}_j}{N}+\frac{1}{2}
	\left(q^{kl}-\frac{\beta^k\beta^l}{N^2}\right)\scr{A}_k\scr{A}_l
		  \right]\partial_i(\nabla^{\mu}D_{\mu}\phi)
		       \right\}, \nonumber\\
 T^{ij} & \equiv & 
  -\frac{2}{\sqrt{q}}\frac{\delta}{\delta q_{ij}}
  \int dtd^3xN\sqrt{q}\scr{L}
  \nonumber\\
 & = & \frac{M^2}{g^2}
  \left\{
   \left(\frac{1}{2}q^{kl}F_{\perp k}F_{\perp l}
    -\frac{1}{4}q^{km}q^{ln}F_{kl}F_{mn}\right)q^{ij}
   -q^{ik}q^{jl}F_{\perp k}F_{\perp l} + q^{ik}q^{jl}q^{mn}F_{km}F_{ln}
   \right.\nonumber\\
 & & \left. -q^{ik}q^{jl}\scr{A}_k\scr{A}_l
   \frac{1}{\sqrt{q}}\partial_m(\sqrt{q}q^{mn}F_{\perp n})\right\}
 \nonumber\\
 & & + \alpha M^2
  \left\{
   \left[ \frac{1}{2}(\nabla^{\mu}D_{\mu}\phi)^2
   \right.\right.\nonumber\\
 & & \left.\left.
    - \left(1-\frac{\beta^k\scr{A}_k}{N}
       +\frac{1}{2}\left(q^{kl}-\frac{\beta^k\beta^l}{N^2}\right)
       \scr{A}_k\scr{A}_l\right)\partial_{\perp}(\nabla^{\mu}D_{\mu}\phi)
    +q^{kl}\scr{A}_k\partial_l(\nabla^{\mu}D_{\mu}\phi)\right]q^{ij}
   \right.\nonumber\\
 & & \left.+q^{ik}q^{jl}\scr{A}_k\scr{A}_l
      \partial_{\perp}(\nabla^{\mu}D_{\mu}\phi)
   -q^{ik}q^{jl}(\scr{A}_k\partial_l+\scr{A}_l\partial_k)
   (\nabla^{\mu}D_{\mu}\phi)	\right\}.
\end{eqnarray}
The Einstein equation is
%
\begin{equation}
 \MP^2G_{\mu\nu}u^{\mu}u^{\nu} = T_{\perp\perp}, \quad
 \MP^2G_{\mu i}u^{\mu} = T_{\perp i}, \quad
 \MP^2G^{ij} = T^{ij},
\end{equation}
where
%
\begin{equation}
 u^{\mu} = \left(\frac{\partial}{\partial t}\right)^{\mu}
  -\beta^i\left(\frac{\partial}{\partial x^i}\right)^{\mu}.
\end{equation}

Just for completeness, the equation of motion for $\scr{A}_i$ is
%
\begin{eqnarray}
 & & \frac{1}{g^2}
  \left\{
   \frac{1}{\sqrt{q}}\partial_{\perp}(\sqrt{q}q^{ij}F_{\perp j})
   -\left[
     \left(q^{ij}-\frac{\beta^i\beta^j}{N^2}\right)\scr{A}_j
     -\frac{\beta^i}{N} \right]\frac{1}{\sqrt{q}}
   \partial_k(\sqrt{q}q^{kl}F_{\perp l}) \right.\nonumber\\
 & & \left.
   + \frac{1}{N\sqrt{q}}\partial_j(N\sqrt{q}q^{ik}q^{jl}F_{kl})
   + \left(
      q^{kj}F_{\perp j}\frac{\partial_k\beta^i}{N}
      - q^{ij}F_{\perp j}\frac{\partial_k\beta^k}{N}\right)\right\}
  \nonumber\\
  & & + \alpha
  \left\{
   \left[
    \left(q^{ij}-\frac{\beta^i\beta^j}{N^2}\right)\scr{A}_j
    -\frac{\beta^i}{N}\right]\partial_{\perp}(\nabla^{\mu}D_{\mu}\phi)
   -q^{ij}\partial_j(\nabla^{\mu}D_{\mu}\phi)\right\}
  = 0.
\end{eqnarray}

\section{Static Post-Newtonian Gravity}
\label{app:ppn}

The PPN parameters $\beta_{\rm PPN}$ and $\gamma_{\rm PPN}$ are defined in the
$1/r$-expansion of a spherically symmetric, static metric in the
isotropic coordinate as 
%
\begin{equation}
 ds^2 = - \left[ 1 - \frac{2m}{r} + \frac{2\beta_{\rm PPN} m^2}{r^2}
	   + O(m^3/r^3) \right] dt^2
 + \left[ 1 + \frac{2\gamma_{\rm PPN} m}{r} + O(m^2/r^2) \right] 
 \left( dr^2 + r^2 d\Omega_2^2 \right). 
\end{equation}
In this coordinate, $\beta_{\rm PPN}$ and $\gamma_{\rm PPN}$, respectively,
measure the amount of non-linearity and the amount of space curvature
produced by a mass. The values in General Relativity is
$\beta_{PPN}=\gamma_{PPN}=1$. Experimental limits on them are 
%
\begin{equation}
 |\beta_{\rm PPN}-1| < 10^{-3}, \quad |\gamma_{\rm PPN}-1| < 10^{-3}.
\end{equation}
In this section we calculate $\beta_{\rm PPN}$ and $\gamma_{\rm PPN}$
for the gauged ghost condensate.

For the isotropic coordinate
%
\begin{equation}
 ds^2 = - N(r)^2dt^2 + B(r)^2 \left( dr^2 + r^2 d\Omega_2^2 \right),
\end{equation}
Einstein tensor is given by
%
\begin{eqnarray}
 G_{\perp\perp} & = & \frac{1}{B^2}
  \left[ \left(\frac{B'}{B}\right)^2 - \frac{4B'}{rB} -
   \frac{2B''}{B}\right],
 \nonumber\\
 G_{\perp r} & = & G_{\perp\theta} = 0, \nonumber\\
 G^{rr} & = & \frac{1}{B^4}
  \left[2\frac{B'}{B}\frac{N'}{N} + \left(\frac{B'}{B}\right)^2
   + \frac{2}{r}\frac{B'}{B} + \frac{2}{r}\frac{N'}{N}\right], \nonumber\\
 G^{\theta\theta} & = & \frac{1}{r^2B^4}
  \left[ \frac{N''}{N}+\frac{N'}{rN}+\frac{B'}{rB}+\frac{B''}{B}
   -\left(\frac{B'}{B}\right)^2\right], \nonumber\\
  G^{r\theta} & = & 0.
\end{eqnarray}
We shall apply the formulae obtained in App.~\ref{app:Tmunu} to this
metric by setting 
%
\begin{equation}
 N = N(r), \quad
  \beta^i = 0, \quad 
  q_{ij}dx^idx^j = B(r)^2 \left( dr^2 + r^2 d\Omega_2^2\right), \quad
  A_idx^i = A(r)dr.
\end{equation}
The result is
%
\begin{eqnarray}
 M^{-2}T_{\perp\perp} & = & \frac{1}{g^2}
  \left[ \frac{1}{2}\left(\frac{F_{\perp}}{B}\right)^2
   -\left(1+\frac{A^2}{2B^2}\right)\frac{(r^2BF_{\perp})'}{r^2B^3}
  \right]
  - \alpha\left[\frac{1}{2}(dA)^2+\frac{A(dA)'}{B^2}\right],
  \nonumber\\
 M^{-2}T_{\perp r} & = & -\alpha\left(1+\frac{A^2}{2B^2}\right)(dA)', 
  \nonumber\\
 M^{-2}T_{\perp\theta} & = & 0,\nonumber\\
 M^{-2}T^{rr} & = & \frac{1}{B^2}
  \left\{-\frac{1}{g^2}
  \left[ \frac{1}{2}\left(\frac{F_{\perp}}{B}\right)^2
   +\frac{A^2(r^2BF_{\perp})'}{r^2B^5}\right]
  +\alpha\left[\frac{1}{2}(dA)^2 - \frac{A(dA)'}{B^2}\right]\right\}, 
  \nonumber\\
 M^{-2}T^{\theta\theta} & = &
  \frac{1}{r^2B^2}\left\{
  \frac{1}{2g^2}\left(\frac{F_{\perp}}{B}\right)^2 
  + \alpha
  \left[ \frac{1}{2}(dA)^2 + \frac{A(dA)'}{B^2}\right]\right\},
  \nonumber\\
 M^{-2}T^{r\theta} & = & 0,
\end{eqnarray}
where
%
\begin{eqnarray}
 F_{\perp} & = & -\frac{1}{N}
  \left[ N\left(1+\frac{A^2}{2B^2}\right)\right]', \nonumber\\
 dA & = & \frac{(r^2NBA)'}{r^2NB^3}. 
\end{eqnarray}

In order to calculate $\beta_{\rm PPN}$ and $\gamma_{\rm PPN}$, we expand
all variables by $1/r$,
%
\begin{eqnarray}
 N(r) & = & 1 + \sum_{n=1}^{\infty}\frac{N_n}{r^n}, \nonumber\\
 B(r) & = & 1 + \sum_{n=1}^{\infty}\frac{B_n}{r^n}, \nonumber\\
 A(r) & = & \sum_{n=0}^{\infty}\frac{A_n}{r^n}.
\end{eqnarray}
Accordingly, we obtain the $1/r$-expansion of the Einstein equation.

The ($\perp r$)-component of the Einstein equation is $\alpha(dA)'=0$,
which with the above $1/r$-expansion implies that $A=0$. In this case 
non-vanishing components of the stress energy tensor are
%
\begin{eqnarray}
 M^{-2}T_{\perp\perp} & = & \frac{1}{g^2}
  \left[ \frac{1}{2}\left(\frac{N'}{NB}\right)^2
   + \left(1+\frac{A^2}{2B^2}\right)
   \frac{1}{r^2B^3}\left(\frac{r^2BN'}{N}\right)'
  \right],  \nonumber\\
 B^2M^{-2}T^{rr} & = & -\frac{1}{g^2}
  \left[ \frac{1}{2}\left(\frac{N'}{NB}\right)^2
   +\frac{A^2}{r^2B^5}\left(\frac{r^2BN'}{N}\right)'
  \right], \nonumber\\
 r^2B^{-2}M^{-2}T^{\theta\theta} & = &
  \frac{1}{2g^2}\left(\frac{N'}{NB}\right)^2. 
\end{eqnarray}
The leading term in the ($rr$)-component of the Einstein equation says
that
%
\begin{equation}
 B_1 = -N_1. \label{eqn:B1}
\end{equation}
With this relation, the leading term in 
$G_{\perp\perp}-B^2G^{rr}=\kappa^2(T_{\perp\perp}-B^2T^{rr})$ is
%
\begin{equation}
 (M^2-2g^2\MP^2)\left(N_2-\frac{N_1^2}{2}\right) = 0. 
\end{equation}
Hence, unless $M^2=2g^2\MP^2$, we obtain
%
\begin{equation}
 N_2 = \frac{N_1^2}{2}. \label{eqn:N2}
\end{equation}

{}From (\ref{eqn:B1}) and (\ref{eqn:N2}) we obtain 
%
\begin{equation}
 \beta_{\rm PPN} = \gamma_{\rm PPN} = 1. 
\end{equation}
Therefore, there is no constraint from the spherically symmetric, static
post-Newtonian gravity. The leading correction to General Relativity
appears in $B_2/N_1^2$. Actually, the leading term in the
($\perp\perp$)-component of the Einstein equation says that 
%
\begin{equation}
 \frac{B_2}{N_1^2} =
  \frac{1}{4}\left(1+\frac{M^2}{2g^2\MP^2}\right),
\end{equation}
and this value of $B_2/N_1^2$ has a deviation from the GR value
$1/4$. However, this deviation is in the post-post Newtonian order and
beyond the current ability of gravity experiments.

\section{Fragmentation of Planar Caustics}
\label{app:fragmentation}

In Sec.~\ref{sec:nonlinear} we have provided both analytical
and numerical evidences showing that there should be no instability
except for the perfectly plane symmetric case. We have also pointed out
that, on the other hand, it is in principle possible to create a caustic
plane if the perfectly planar symmetry is assumed. However, this
situation is unlikely to happen in generic situations since the perfect
plane symmetry is too ideal an assumption. Indeed, small fluctuations on
top of the plane symmetric collapse should grow and the layer should be 
broken into pieces before a planar caustics occurs. 
After the fragmentation, a lower-dimensional caustics does not occur
since, as discussed in subsection~\ref{subsec:Caustics}, for
codimension $2$ or higher it would cost infinite amount of work to
compress a finite-volume, negative $r^sE_r$ ($s=1,2$) region to an
infinitesimal volume or thickness.

In this appendix we perform a perturbative analysis supporting this
picture of fragmentation. In particular, we analyze inhomogeneous linear
perturbation on top of a perfectly planer caustic solution and show that
inhomogeneous perturbation grows faster than the background caustics.

In a fixed flat spacetime background $ds^2=dt^2-dx^2-dz^2$, we consider
a configuration depending on two space coordinates $x$ and $z$ and the
time $t$:
%
\begin{equation}
 A_idx^i = A_x(t,x,z)dx + A_z(t,x,z)dz.
\end{equation}
The equation of motion is then
%
\begin{eqnarray}
 \dot{E}_i & = & A_i\partial^jE_j -\partial^jF_{ij}
  + \alpha g^2\partial_i\partial^jA_j, \nonumber\\
 E_i & \equiv & \dot{A}_i - A^j\partial_iA_j,
\end{eqnarray}
where $i$ and $j$ run over $x$ and $z$. We consider a planar scaling
solution with inhomogeneous linear perturbation of the form 
%
\begin{equation}
 A_z = \frac{z}{-t}\left[ 1 + f(t)e^{ikx}\right], \quad 
  A_x =  g(t)e^{ikx},
\end{equation}
where we consider $f$ and $g$ as small perturbations. Since we are
interested in the behavior near the caustic plane $z=0$, we have
truncated the Taylor expansion w.r.t. $z$ at the lowest order. In 
the first order in $f$ and $g$, the equation of motion becomes 
%
\begin{eqnarray}
 \ddot{f} - \frac{\dot{f}}{-t}
  + \left[k^2-\frac{k^2z^2+1}{(-t)^2}\right]f
  + ik\left[\frac{\dot{g}}{-t}+\frac{g}{(-t)^2}\right] & = & 0,
  \nonumber\\
 \ddot{g} + \frac{2\dot{g}}{-t}+ \frac{2g}{(-t)^2}
  +ikz^2\frac{\dot{f}}{-t}+ik\left[\frac{2z^2}{(-t)^2}-1\right]f
  & = & 0,
\end{eqnarray}
By eliminating $g$ we obtain the third order equation for $f$ as 
%
\begin{equation}
 \dddot{f} - \frac{\ddot{f}}{-t} 
  + \left[k^2-\frac{2}{(-t)^2}\right]\dot{f}
  -\left[k^2+\frac{2}{(-t)^2}\right]\frac{f}{-t}
  = 0.
\end{equation}
The general solution to this equation is
%
\begin{equation}
 f = \frac{C_1}{-t} + C_2\left[k\cos(kt)+\frac{1}{-t}\sin(kt)\right]
  + C_3\left[k\sin(kt)-\frac{1}{-t}\cos(kt)\right],
\end{equation}
where $C_1$, $C_2$ and $C_3$ are constants. Hence, the linear
perturbation is unstable and grows faster than the background.


\section{A Simple Way to Calculate the Energy Loss in \v{C}erenkov Radiation}
\label{sec:simpleway}

Consider a Lagrangian in the presence of an external source
\begin{equation}
{\cal L}_{\rm Total} = {\cal L}_{\phi}(\phi,\, \partial \phi) + {\cal L}_S(\phi,\, S),
\end{equation}
where ${\cal L}_S$ represent the source term, {\it e.g.} ${\cal L}_S = S\phi$.
If the action without the source ${\cal S}_{\phi} = \int d^4 x {\cal L}_{\phi}$ is invariant under the transformation $\phi \to \phi'= \phi + \delta \phi$, there is an associated conserved Noether current
\begin{equation}
J_\phi^\mu = \frac{\delta {\cal L}_{\phi}}{\delta (\partial_\mu \phi)} \delta \phi -V^\mu,
\end{equation}
with $V^\mu$ given by
\begin{equation}
\delta {\cal L}_{\phi} = \partial_\mu V^\mu.
\end{equation}
The presence of the (fixed) external source ``breaks'' the invariance of the transformation, so the current is no longer conserved:
\begin{equation}
\partial_\mu J_\phi^\mu = \delta {\cal L}_S.
\end{equation}
Integrating over space, we obtain
\begin{equation}
\int d^3 x \, \delta {\cal L}_S = \int d^3 x \,\partial_\mu J_\phi^\mu = \partial_0 Q_\phi + \oint J^i d S_i .
\end{equation}
The right hand side is just the total charge flowing into the $\phi$ field and out to infinity, so it must be compensated by the loss of the total charge of the source,
$-dQ_S/dt$:
\begin{equation}
\frac{dQ_S}{dt} = -\int d^3 x \, \delta {\cal L}_S.
\end{equation}
For energy, the corresponding symmetry transformation is time translation, so $\delta \phi = \dot{\phi}$. The energy loss for a source ${\cal L}_S=S\phi$ is then simply given by
\begin{equation}
\label{eq:energy_change}
\frac{dE_S}{dt} =- \int d^3 x\, S\, \dot{\phi}.
\end{equation}
In the linearized approximation, $\dot{\phi}$ can be easily solved in terms of a given source in momentum space. It is then straightforward to compute the energy loss using \Eq{energy_change}. It applies to both the \v{C}erenkov radiation when a source moving in a medium faster than the sound speed, and multipole radiations due to accelerations. As an example, we derive the energy loss from the Goldstone boson radiation in a binary system. The results are used in sec.~\ref{sec:Cerenkov} to derive bounds on the scale of the Lorentz violation from the binary pulsars.

We consider a system of binary pulsars where two pulsars of equal mass $M_0$ move in a circular orbit separated by a distance $2r_0$. We ignore the velocity of the center of the system for this calculation.

In the $\delta$-function approximation, the source is given by
\begin{eqnarray}
\rho (x, t) &=& M_0 \delta^3\left(\vec{x} -\vec{r}(t)\right) + M_0 \delta^3 \left(\vec{x} +\vec{r}(t)\right), \\
\vec{r} (t) &=& (r_0 \cos \omega_0 t,\, r_0 \sin \omega_0 t,\, 0),
\end{eqnarray}
where the angular frequency $\omega_0$ is related to the orbital velocity
$v_0$ by $v_0= \omega_0 r_0$. Its Fourier transform is
\begin{equation}
\tilde{\rho} (\omega, k) = 2M_0\sum_{n=-\infty}^{\infty} e^{i\, 2n (\theta_k-\frac{\pi}{2})} J_{2n}(k_{\parallel} r_0) \, 2\pi \delta(\omega -2n\omega_0),
\end{equation}
where $k_{\parallel}= \sqrt{k_x^2+k_y^2}$, $\tan \theta_k = k_y/k_x$, and we
have used the property of the Bessel functions, $J_m(-x)=(-1)^m J_m(x)$.
A form factor $f(k)$ can also be included to represent the finite size effect. 
Applying the simple way of calculating the energy loss described above and using Eqs.(\ref{eq:L_S})--(\ref{eq:response}), we obtain the time-averaged energy loss rate
\begin{equation}
\label{eq:Bessel_sum}
\frac{dE_S}{dt} = - \frac{\alpha M_0^2 M^2}{4\pi M_{\rm Pl}^4 c_s} \sum_{n=-\infty}^{\infty} \int_{0}^{1} d(\cos \theta) \left(\frac{2n\omega_0}{c_s}\right)^2 J_{2n}\left(\frac{2n\omega_0 r_0 \sin \theta}{c_s}\right)^2 \left|f\left(\frac{2n\omega_0}{c_s}\right)^2\right|.
\end{equation}

Now we can consider two different limits. First, for $v_0= \omega_0 r_0 \ll c_s$ we can use the approximation
\begin{equation}
J_m (x) \simeq \frac{1}{m !} \left( \frac{x}{2}\right)^m, \; \mbox{ for small } x.
\end{equation}
The sum is dominated by the smallest $|n|$, $n=\pm 1$ which corresponds to quadrupole radiation. There is no form factor suppression ($f(k)\simeq 1$) because the inverse of the momentum is much bigger than the size of the system. The resulting energy loss by the quadrupole radiation is
\begin{equation}
\label{eq:case1}
\frac{d E_S}{dt} = - \frac{4\alpha M_0^2 M^2 v_0^6}{15\pi M_{\rm Pl}^4 r_0^2 c_s^7} = -\frac{2^{12}\pi \alpha M^2 v_0^{10}}{15\, c_s^7},
\end{equation}
where in the last equality we have used the relation $G_N M_0/r_0= 4 v_0^2$.

In the opposite limit, $c_s \ll v_0$, we need to use the asymptotic approximation of the Bessel functions,
\begin{equation}
J_m(x) \simeq \sqrt{\frac{2}{\pi x}} \cos \left[ x-\left(m+\frac{1}{2}\right)\frac{\pi}{2}\right],\;\;\; \mbox{ for } x \gg 1.
\end{equation}
{}From \Eq{Bessel_sum} the energy loss is given approximately by
\begin{equation}
\label{eq:case2}
\frac{dE_S}{dt} \simeq -\frac{\alpha M^2 M_0^2}{4\pi M_{\rm Pl}^4 c_s r_0^2} \sum_{n=1}^{\infty} \frac{2n v_0}{c_s} \left| f\left(\frac{2n v_0}{c_s r_0}\right)^2 \right|.
\end{equation}

\section{A Null Test for a Spin-Dependent Force Law}
\label{ringofspin}

It is well known that in electromagnetism, the magnetic field outside of a divergence-free spin configuration vanishes.  Similarly for (ungauged) ghost condensation in the non-relativistic limit, the coupling between spin and ghostone boson field $\pi$ takes the form \cite{Arkani-Hamed:2004ar}
\beq[generalizationpiint]
\mathcal{L}_{\rm int} = \frac{M}{F} \vec{s} \cdot \grad \pi,
\eeq
and doing an integration by parts, we see that $\pi$ is not sourced by a divergence-free spin configuration.  However, the potential in \Eq{novelocitypotential} is not of the form of \Eq{generalizationpiint}.  In particular, the transverse modes of $\vec{A}_i$ couple directly to $\vec{s}$ and not to its divergence.

One interesting divergence-free spin configuration is a ring of spin.  This could be established with a toroidal electromagnet, and because the magnetic field is zero outside of the torus, this allows a null test for the gauged ghost condensate spin potential.   Consider the following setup, where a ring of radius $R$ with net spin $S_1$ is separated a distance $r$ away from a point spin $S_2$.  
\beq
\begin{tabular}{c}
\includegraphics{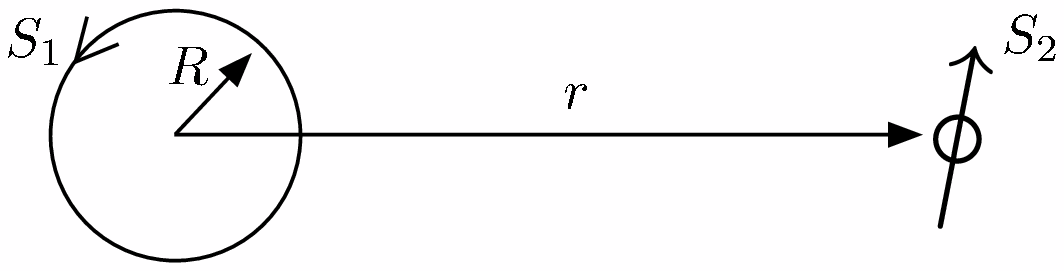}
\end{tabular}
\eeq
It is straightforward to calculate the potential from \Eq{novelocitypotential}:
\beq
V(r) = \frac{M^2 g^2}{F^2}\frac{S_1 S_{2y}}{8 \pi r}  \frac{R}{r} + \scr{O}(R^3/r^3),
\eeq
where $S_{2y}$ is the projection of $S_2$ in the vertical direction.  Note that this potential does not depend on $\alpha$, because $\alpha$ only controls the coupling of the longitudinal mode of $\vec{A}$.  Though this potential goes as $1/r^2$, there is in principle no magnetic field leakage from the ring $S_1$, so it should be easier to design a null experiment to look for this long-range spin-spin potential.

\end{document}